\documentclass[lettersize,journal]{IEEEtran}
\usepackage{amsmath,amsfonts}
\usepackage{algorithm}
\usepackage{array}
\usepackage{textcomp}
\usepackage{stfloats}
\usepackage{url}
\usepackage{verbatim}
\usepackage{graphicx}
\usepackage{cite}
\usepackage{booktabs}
\usepackage{multirow}
\usepackage{makecell}
\usepackage{subfigure}
\usepackage[misc]{ifsym}
\usepackage{listings}
\usepackage{color, xcolor}
\usepackage{algpseudocode}
\usepackage{tcolorbox}
\usepackage{enumitem}

\usepackage{amssymb}
\usepackage{graphicx}
\usepackage{algorithmicx}
\usepackage{balance}

\begin{document}

\title{Towards In-Depth Root Cause Localization \\ for Microservices with Multi-Agent Recursion-of-Thought}

\author{
	\IEEEauthorblockN{Lingzhe Zhang, Tong Jia\IEEEauthorrefmark{1}, Kangjin Wang, Chiming Duan, Minghua He,\\ Rongqian Wang, Xi Peng, Meiling Wang, Gong Zhang, Renhai Chen and Ying Li\IEEEauthorrefmark{1},~\IEEEmembership{Member,~IEEE}}
	\thanks{Lingzhe Zhang, Tong Jia, Kangjin Wang, Chiming Duan, Minghua He, and Ying Li are with Peking University, Beijing, China.}
	\thanks{Rongqian Wang, Xi Peng, Meiling Wang, Gong Zhang, and Renhai Chen are with Huawei Theory Lab, China.}
	\thanks{Email: \{zhang.lingzhe, duanchiming, hemh2120\}@stu.pku.edu.cn, \{jia.tong, wangkangjin, li.ying\}@pku.edu.cn and \{wangrongqian2, pancy.pengxi, wangmeiling17, nicholas.zhang, chenrenhai\}@huawei.com}
	\thanks{* Corresponding author: Tong Jia, e-mail: (jia.tong@pku.edu.cn); Ying Li, e-mail: (li.ying@pku.edu.cn)}
	\thanks{Manuscript received xxx, 2025; revised xxx.}
}

\markboth{Towards In-Depth Root Cause Localization for Microservices with Multi-Agent Recursion-of-Thought}%
{Shell \MakeLowercase{\textit{et al.}}: A Sample Article Using IEEEtran.cls for IEEE Journals}


\maketitle

\begin{abstract}
	As modern microservice systems grow increasingly complex due to dynamic interactions and evolving runtime environments, they experience failures with rising frequency. Ensuring system reliability therefore critically depends on accurate root cause localization (RCL). While numerous traditional machine learning and deep learning approaches have been explored for this task, they often suffer from limited interpretability and poor transferability across deployments. More recently, large language model (LLM)-based methods have been proposed to address these issues. However, existing LLM-based approaches still face two fundamental limitations: context explosion, which dilutes critical evidence and degrades localization accuracy, and serial reasoning structures, which hinder deep causal exploration and impair inference efficiency. In this paper, we conduct a comprehensive study of both how human SREs perform root cause localization in practice and why existing LLM-based methods fall short. Motivated by these findings, we introduce RCLAgent, an in-depth root cause localization framework for microservice systems that realizes multi-agent recursion-of-thought with parallel reasoning. RCLAgent decomposes the diagnostic process along the trace graph by assigning each span to a Dedicated Agent and organizing agents recursively and in parallel according to the graph topology, with the final diagnosis obtained by synthesizing the Root-Level Diagnosis Report and the Global Evidence Graph. Extensive experiments on multiple public benchmarks demonstrate that RCLAgent consistently outperforms state-of-the-art methods in both localization accuracy and inference efficiency.
\end{abstract}


\section{Introduction}

Modern microservice systems have become increasingly complex due to dynamic interactions and evolving runtime environments~\cite{zhou2018fault, zhang2025survey}. These systems often consist of hundreds or even thousands of fine-grained, interdependent subsystems, where issues in any one component can easily lead to performance problems at the top level~\cite{mendoncca2019developing, waseem2021design, zhang2024towards, zhang2024time, kang2022separation}. Therefore, to ensure system reliability, it is crucial to localize the root cause of these issues in a timely manner~\cite{zhang2024multivariate, zhang2024reducing}.

However, root cause localization in microservice systems is intrinsically difficult due to the dense and dynamic interdependencies among subsystems~\cite{wang2023interdependent, zhang2024failure, yu2024survey, sun2025interpretable, zhu2024hemirca, xie2024microservice, wang2024kgroot, pham2025rcaeval, zheng2024lemma}. A single user request often propagates through a long and branching invocation path, spanning services, instances, hosts, and heterogeneous interactions such as RPC calls and database operations. These invocation patterns evolve continuously as deployments change and workloads fluctuate, resulting in highly dynamic and heterogeneous execution contexts. Such complexity makes it non-trivial to determine where an anomaly truly originates.

To cope with this challenge, existing research has primarily followed two paradigms. Graph-based approaches model inter-service relationships using dependency or causality graphs, often combined with statistical or learning-based scoring mechanisms~\cite{lin2018microscope, li2022causal, yu2021microrank, yu2023tracerank, zhang2022crisp, zhang2024trace, lin2024root, yu2023nezha}. While these methods provide explicit and interpretable reasoning paths, they depend on rigid graph abstractions that are difficult to adapt across heterogeneous platforms or evolving system topologies. In contrast, deep learning-based techniques learn temporal, spatial, and structural patterns directly from metrics, logs, and traces~\cite{gan2019seer, yang2022micromilts, cai2021modelcoder, ding2023tracediag, wang2023root, chen2021trace, ren2023grace, li2022actionable, lee2023eadro}. Although effective, such models typically operate as black boxes, offering limited interpretability and poor transferability across deployments. Consequently, both paradigms remain constrained by weak cross-platform generalization and limited adaptability to changing system behaviors.

Recent advances in large language models (LLMs) open up a new avenue for overcoming these limitations, and an increasing number of studies have begun to explore their use in root cause analysis~\cite{zhang2024mabc, wang2024rcagent, pei2025flow, li2025coca, wang2025tamo, ren2025multi, roy2024exploring, shi2024enhancing, xie2024cloud, han2024potential, zhang2025thinkfl, zhang2025scalalog, zhang2025agentfm, zhang2024automated, shan2024face, xu2025openrca}. Existing efforts can be roughly grouped into three strands: (i) early explorations that demonstrate the feasibility and potential of LLMs for diagnostic reasoning~\cite{sarda2024leveraging, roy2024exploring, shi2024enhancing, xie2024cloud, han2024potential, zhang2025agentfm}; (ii) structured solutions that organize reasoning through multi-agent architectures~\cite{zhang2024mabc, wang2024rcagent, pei2025flow, wang2025tamo, ren2025multi}; and (iii) knowledge-enhanced approaches that ground LLMs with external information via techniques such as retrieval-augmented generation (RAG)~\cite{zhang2024automated, sarda2024leveraging, shan2024face, li2025coca}.

Although existing LLM-based root cause localization methods have demonstrated promising results, they still face fundamental practical challenges when applied to real-world microservice systems:

\begin{itemize}
	\item \textbf{Context Explosion Dilutes Evidence and Limits Accuracy.}  
	Existing approaches either ingest large volumes of runtime data into a single prompt, or adopt step-by-step paradigms such as ReAct. Even in multi-agent designs, each agent is typically provided with an ever-growing, accumulated context. As microservice executions naturally form deep and branching trace graphs, this design quickly leads to \emph{context explosion}, often exceeding the token budget of LLMs. When overflow occurs, critical causal signals are truncated; even when it does not, the model is forced to reason under severe cognitive load, causing subtle but decisive dependencies to be overlooked. Since root cause localization inherently requires drilling down to the deepest causal factor, such flattened and overloaded contexts dilute evidence and bias the model toward coarse hypotheses, ultimately degrading diagnostic accuracy.
	
	\item \textbf{Serial Reasoning Structures Impede Deep Exploration and Limit Efficiency.}  
	Most existing frameworks follow a strictly sequential workflow, typically in the form of \emph{observe $\rightarrow$ think $\rightarrow$ call tool $\rightarrow$ think $\rightarrow$ \dots}. Under this paradigm, all reasoning steps are executed in series. Each additional exploration step incurs higher latency and a longer context, implicitly making deeper investigation increasingly expensive. As a result, LLMs tend to prematurely converge on intermediate hypotheses before reaching the deepest causal factors, leading to \emph{shallow reasoning} and reduced accuracy. Meanwhile, the strictly serial execution also causes end-to-end latency to grow almost linearly with system complexity: the more services and spans involved in a trace, the longer the diagnosis takes. This serialized reasoning paradigm is fundamentally misaligned with the inherently parallel structure of microservice systems, making timely diagnosis difficult in large-scale deployments and limiting efficiency.
\end{itemize}

Recognizing this gap, we first perform a lightweight empirical analysis to understand how SREs localize root causes in practice and why existing LLM-based methods fall short. By synthesizing insights from prior studies and professional SRE experience, we distill three recurring characteristics of manual root cause analysis: \emph{recursiveness}, \emph{multi-dimensional expansion}, and \emph{cross-modal reasoning}. We then conduct an in-depth experimental analysis of representative LLM-based approaches, revealing that their limitations stem primarily from \emph{context explosion} and \emph{serial reasoning}.

Building on these insights, we introduce \textbf{RCLAgent}\footnote{Code Repository: \url{https://github.com/LLM4AIOps/RCLAgent-V2}}, an in-depth root cause localization framework for microservice systems that realizes multi-agent recursion-of-thought with parallel reasoning.

To address the first challenge, RCLAgent decomposes the diagnostic process along the structure of the trace graph. Each span is assigned a dedicated agent, and agents are organized recursively following the graph topology. Every agent operates within a strictly bounded context: it performs \emph{self-state verification} by invoking log and metric analysis tools on its own span, and consolidates only the \emph{downstream evidences} provided by its child agents. Rather than propagating raw observations upward, each agent emits a compact, structured hypothesis with a concise rationale for its parent. In parallel, all local hypotheses are retained in a \emph{global evidence graph}, which is jointly analyzed at the root level. This hierarchical abstraction prevents unbounded context growth and suppresses evidence dilution, allowing decisive causal signals to remain salient while preserving deep causal chains. As a result, RCLAgent can drill down to the true root cause without overwhelming the model, improving diagnostic accuracy.

To address the second challenge, RCLAgent exploits the intrinsic parallelism of multi-agent recursion-of-thought. Sub-agents corresponding to independent branches of the trace graph reason concurrently, making deeper and broader exploration no longer incur prohibitive sequential cost. By decoupling reasoning depth from wall-clock latency, RCLAgent removes the implicit pressure to "stop early", which in existing serial pipelines often forces LLMs to settle on intermediate hypotheses before reaching the deepest causal factors. To ensure scalability, RCLAgent further introduces an \emph{Agents Pool} that caps the number of concurrently active agents, enabling controlled parallelism. Together, these mechanisms not only reduce end-to-end latency but also enable sustained deep exploration, effectively mitigating shallow reasoning while improving diagnosis efficiency.

We conduct experiments on the AIOPS 2022~\cite{aiops2022championship}, Augmented-TrainTicket~\cite{yu2023nezha}, and RCAEval~\cite{pham2025rcaeval} datasets to evaluate RCLAgent. The results show that, in terms of accuracy, RCLAgent outperforms both other LLM-based and non-LLM-based root cause localization methods, surpassing the second-best approach by approximately 7.51\%. In terms of inference efficiency, RCLAgent achieves more than 1.75× speedup over existing LLM-based methods. In summary, the key contributions of this work are as follows:

\begin{itemize}
	\item We conduct a comprehensive study of how SREs localize root causes, revealing that human analysis is inherently recursive, multi-dimensional, and cross-modal, while existing LLM-based methods fall short due to context explosion and serial reasoning.
	\item Inspired by these insights, we propose RCLAgent, an in-depth root cause localization framework for microservice systems that realizes multi-agent recursion-of-thought with parallel reasoning.
	\item We evaluate RCLAgent on multiple public datasets, demonstrating its effectiveness. Experimental results show that RCLAgent achieves superior performance in both accuracy and efficiency.
\end{itemize}

\section{Background}

In this section, we present the essential background of this paper, including what is a failure in microservice systems, the formal definition of the root cause localization problem, an overview of traditional root cause localization methods, and an introduction to distributed tracing as the primary data source for root cause localization.

\begin{figure*}[ht]
	\centering
	\includegraphics[width=1\linewidth]{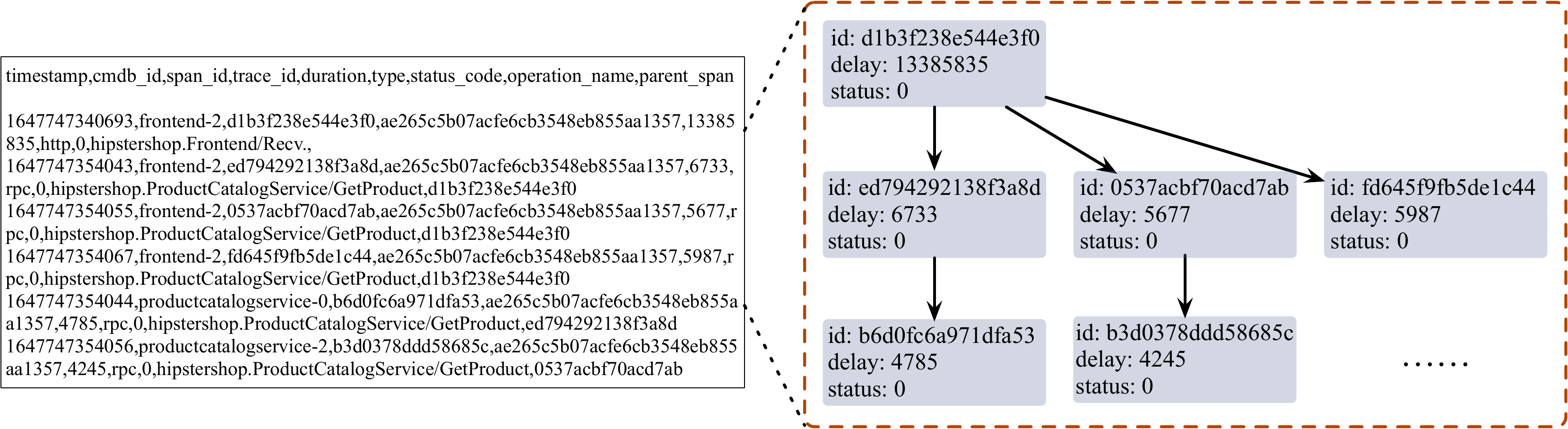}
	\caption{Example of Trace Log}
	\label{fig: tracing}
\end{figure*}

\subsection{Failure in Microservice Systems}
\label{sec:failure}

In large-scale microservice systems, \emph{failures} refer to the actual inability of a service or component to correctly or efficiently perform its intended functionality. Such failures rarely manifest directly; instead, they are revealed through observable \emph{anomalies} in system behavior. As discussed in prior work~\cite{soldani2022anomaly}, while failures denote the underlying faults, anomalies correspond to their observable symptoms, such as increased response latency, reduced throughput, or repeated error events recorded in logs.

At the system level, these anomalies are reflected in observability signals, including abnormal spikes in service metrics (e.g., error rates, CPU usage, or memory consumption), suspicious log patterns (e.g., repeated timeout or exception messages), and irregular distributed traces (e.g., incomplete spans or delayed downstream calls). Individually, such signals provide only partial and indirect evidence of the underlying failure and rarely identify the root cause on their own.

In this work, we diagnose failures by analyzing their observable anomalous manifestations. Specifically, we treat anomalies as deviations in observability signals that are detected by SRE monitoring and alerting mechanisms (e.g., SLO violations) within a short diagnostic window (typically one minute). An anomaly is considered to be present when one or more of the following conditions are observed:
\begin{itemize}
	\item \textbf{Correctness or availability anomalies}: a surge in HTTP $5xx$ error rates, increased request timeouts, or sudden service unavailability;
	\item \textbf{Performance anomalies}: tail latency (e.g., $p95/p99$) exceeding the SLO budget, or a sharp throughput drop under stable load;
	\item \textbf{Resource-related anomalies}: abnormal spikes in CPU or memory usage, prolonged GC pauses, elevated I/O wait, or connection pool exhaustion correlated with request-level anomalies.
\end{itemize}

These criteria provide an intuitive operational characterization of failures, rather than the strict failure-identification rule used in our experiments. In the evaluation, a failure episode is formally defined at the request level as an abnormal request whose latency exceeds 100 times the normal average, as described later in Section~\ref{sec:trace-failure}. Therefore, the SLO-oriented anomaly description is used for background motivation, while the 100$\times$ latency rule is used for precise failure identification in our experimental setting.

We focus on application-layer \emph{correctness} and \emph{performance} failures in microservice systems. We explicitly \underline{exclude} security incidents, user-intent or query-quality issues, front-end-only rendering problems, and fine-grained source-level bug localization. Our target is \emph{component-level} localization, identifying faulty entities such as services, pods, or hosts.

\subsection{Root Cause Localization}

Root cause localization is a fundamental task in system failure diagnosis, focusing on pinpointing the concrete system component that gives rise to abnormal runtime behavior. Unlike failure categorization—such as distinguishing between resource saturation and network congestion—root cause localization aims to identify \emph{where} the failure originates. Accurate localization is crucial for enabling targeted remediation and reducing mean time to recovery.

In this work, we define the \emph{root cause} as the minimal system component that directly triggers the observed failure. Depending on the deployment and observability granularity, a root cause may manifest at different levels, including the service, pod, or node level. At the service level, the root cause corresponds to a microservice exhibiting anomalous behavior (e.g., \texttt{CheckoutService}). At the pod level, it refers to a specific instance (e.g., \texttt{checkoutservice-0}) that behaves abnormally. At the node level, the root cause is the physical machine (e.g., \texttt{node-0}) hosting the faulty components.

For each failure episode, we assume that the root cause is unique and occurs at exactly one of these levels. Identifying the failure at its true granularity is essential for effective mitigation. For instance, if elevated request latency is caused by overload on a single service instance \texttt{checkoutservice-0}, then that pod constitutes the root cause, even though the enclosing service and hosting node may also exhibit secondary symptoms.

\begin{equation} 
	\mathcal{R} = \arg\!\!\max_{\{k_1, \dots, k_m\} \subseteq \mathcal{K}} \Big( p(k_1), p(k_2), \dots, p(k_m) \Big)
	\label{eq: fl-def}
\end{equation}

Formally, as defined in Equation~\ref{eq: fl-def}, root cause localization seeks to identify and rank a set of candidate components $\{k_1, \dots, k_m\} \subseteq \mathcal{K}$ according to their likelihood of being responsible for observed anomalies. Each candidate $k_i$ is evaluated independently using a scoring function $p(k_i)$, and the resulting ranked list $\mathcal{R}$ orders candidates from most to least likely root causes. This formulation abstracts away the processing of individual anomalous requests while capturing the relative plausibility of each candidate.

\subsection{Distributed Tracing}
\label{sec:trace-failure}

To support fine-grained fault diagnosis in software systems, distributed tracing has been widely adopted in industrial environments, becoming an integral part of modern software infrastructures~\cite{wang2022characterizing, yang2022capturing, shen2023network}. A distributed trace provides a detailed execution record of a request as it propagates through the system, capturing timing, dependencies, and performance characteristics. Each trace consists of multiple structured log entries, known as spans, which document individual operations along the request's execution path.

As illustrated in Figure~\ref{fig: tracing}, a complete trace represents the end-to-end journey of a request, detailing every intermediate operation and its corresponding latency. Each span records crucial information such as the service name, operations, timestamps, and causal relationships between operations. By analyzing the timing and dependencies of spans, distributed tracing enables precise performance monitoring and facilitates anomaly detection in large-scale distributed systems. This structured representation is particularly valuable for diagnosing latency issues, identifying service bottlenecks, and uncovering failure propagation patterns across microservices.

Each trace contains an entry span that represents the overall execution status of the request. In this paper, we define a request as having an issue if its entry span exhibits an excessively high execution latency—specifically, exceeding 100 times the normal average latency. In such cases, root cause localization is necessary to identify the underlying source of the anomaly.

\section{Empirical Study}

\begin{figure*}[htbp]
	\centering
	\includegraphics[width=1\linewidth]{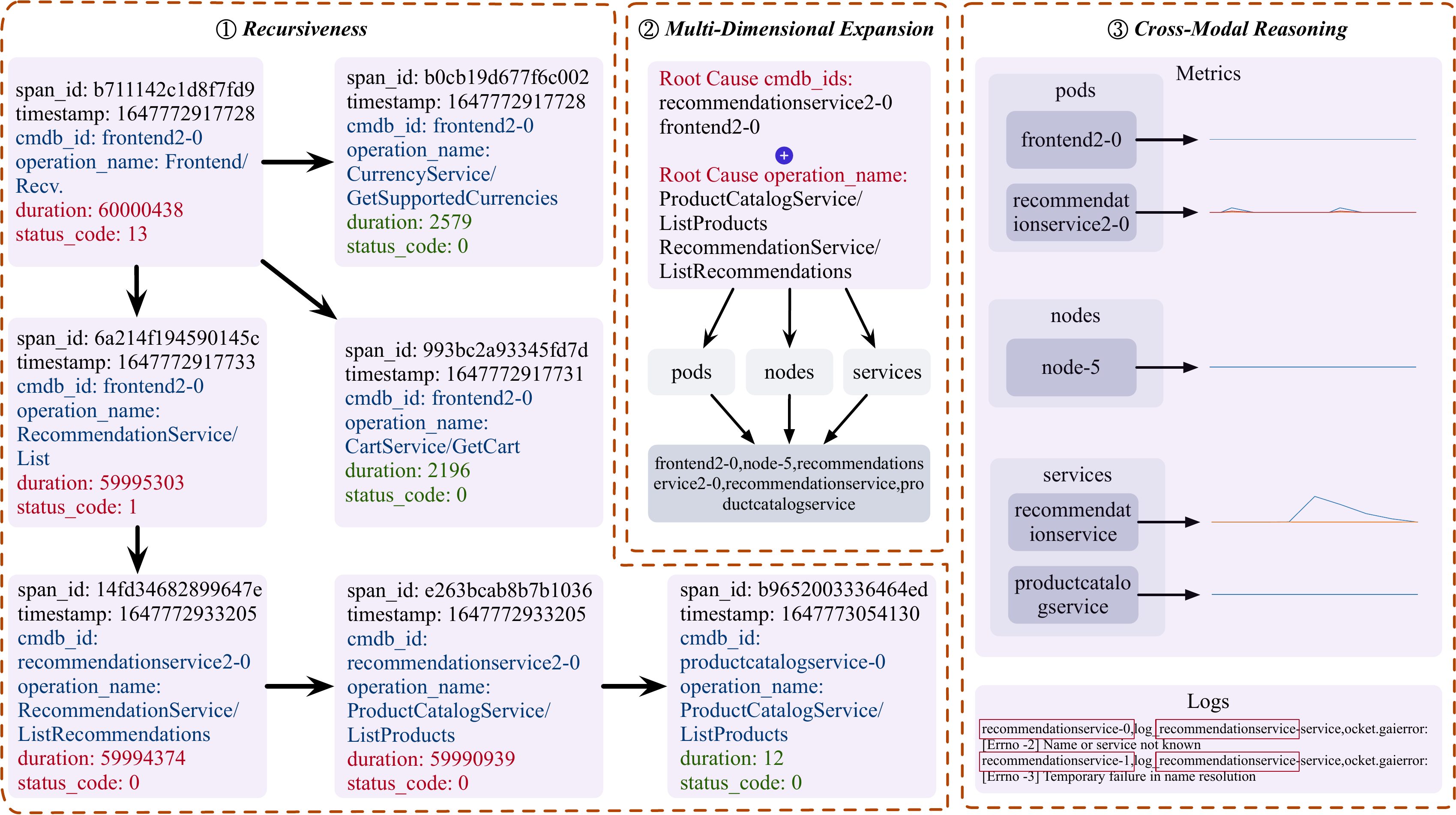}
	\caption{An Illustrative Example of The Manual Root Cause Localization Process}
	\label{fig:empirical-study-example}
\end{figure*}

In this section, we conduct a comprehensive empirical study to explore how an effective and practical root cause localization method should be constructed for microservice systems. To systematically analyze this process, we investigate the following two research questions:

\begin{itemize}
	\item \textbf{RQ1:} How do SREs localize root causes in real-world practice?
	\item \textbf{RQ2:} Why do existing LLM-based methods fall short in this task?
\end{itemize}

\subsection{RQ1: How SREs Perform Root Cause Localization}

We synthesize insights from prior studies to derive an initial characterization of the root cause localization process~\cite{zhang2026agentic, zhang2025thinkfl, pei2025flow, ren2025multi, xiang2025simplifying, yu2023nezha, ding2023tracediag}, and then conduct semi-structured interviews with a total of 15 professional developers and SREs from Peking University and the Huawei Theory Lab to refine and validate this framework. Through this process, we identify three recurring characteristics of root cause analysis.

\textbf{Recursiveness.} SREs typically start from an abnormal request and traverse its trace graph to identify anomalous spans. When a suspicious operation is found, they recursively inspect its downstream dependencies, progressively drilling down along the call chain until the deepest abnormal component is isolated. This top-down, span-by-span refinement enables analysts to efficiently narrow the search space from a complex execution graph to a small set of root causes.

For example, in Figure~\ref{fig:empirical-study-example}, an abnormal request issued by \texttt{frontend2-0} times out and propagates errors along its call chain. By inspecting the trace, SREs first identify that only the \texttt{RecommendationService} branch exhibits abnormal latency. They then recursively follow this branch into its downstream spans, observing repeated timeouts on \texttt{recommendationservice2-0} and its callee operations, while sibling branches remain normal. Through this recursive descent, the vast execution graph is reduced to a small set of suspicious components (e.g., \texttt{frontend2-0} and \texttt{recommendationservice2-0} and their associated operations), shrinking the search space for subsequent diagnosis.

\textbf{Multi-Dimensional Expansion.} While recursion localizes suspicious spans, SREs rarely confine their analysis to a single dimension. Instead, they systematically expand the search scope across multiple axes, correlating anomalous spans with services, pods, and underlying physical nodes. This expansion allows them to reason about failures at different abstraction levels and to uncover latent causes that may not be visible within a single trace path.

In the example shown in Figure~\ref{fig:empirical-study-example}, this expansion is manifested by lifting suspicious trace-level entities to higher-level abstractions. Starting from the anomalous cmdb\_ids and operations, SREs project these signals onto multiple dimensions: pods (\texttt{recommendationservice2-0}, \texttt{frontend2-0}), services (\texttt{recommendationservice}, \texttt{productcatalogservice}), and physical nodes (e.g., \texttt{node-5}). By correlating anomalies across these axes, analysts can reason about whether the failure originates from a specific instance, a service-wide malfunction, or an underlying infrastructure issue, thereby exposing latent causes that are invisible from a single trace path.

\textbf{Cross-Modal Reasoning.} Finally, SREs integrate heterogeneous data modalities—most notably metrics and logs—to distinguish the true root cause from coincidental anomalies. By examining temporal patterns and cross-component correlations in metrics, they identify the component exhibiting the earliest deviation, and then verify this hypothesis through log evidence. Such cross-modal fusion is essential for disambiguating causal faults from merely propagated symptoms.

In Figure~\ref{fig:empirical-study-example}, metric inspection shows that the frontend pod (\texttt{frontend2-0}), the product catalog service (\texttt{productcatalogservice}), and the hosting node (\texttt{node-5}) exhibit no significant fluctuations. In contrast, the recommendation component—at both the pod level (\texttt{recommendationservice2-0}) and the service level (\texttt{recommendationservice})—displays notable variations, with the service-level metrics demonstrating more pronounced and earlier deviations. The anomalies observed at the pod level are propagated from the service itself. Moreover, log records from two replicas of the recommendation service (\texttt{recommendationservice-0} and \texttt{recommendationservice-1}) consistently flag the service as abnormal, effectively ruling out a pod-specific fault. Through this cross-modal reasoning process, the root cause is ultimately identified as the recommendation service, which aligns with the ground truth—network packet corruption in the Kubernetes container.

\begin{center}
	\begin{tcolorbox}[colback=gray!10,
		colframe=black,
		width=\linewidth,
		arc=1mm, auto outer arc,
		boxrule=0.5pt,
		top=2pt, 
		bottom=2pt, 
		left=2pt,
		right=2pt
		]
		\textbf{Summary.} Manual root cause analysis by SREs exhibits three core characteristics: \emph{recursiveness}, where analysts iteratively drill down along the execution path to uncover deeper causal factors; \emph{multi-dimensional expansion}, where the search is broadened across pods, services, and nodes to enumerate all plausible causes; and \emph{cross-modal reasoning}, where hypotheses derived from traces are validated and disambiguated using metrics and logs.
	\end{tcolorbox}
\end{center}

\subsection{RQ2: Why LLM-based Approaches Struggle with RCL}

To analyze the limitations of existing LLM-based methods in root cause localization, we adopt the CoT baseline used in our evaluation, which is implemented as a minimal single-agent ReAct-style CoT diagnostic baseline. This baseline follows a \emph{thought $\rightarrow$ action $\rightarrow$ observation $\rightarrow$ thought $\rightarrow$ \dots} loop, where a single LLM reasons step by step and can invoke trace, log, and metric tools to collect evidence. It does not include additional mechanisms such as multi-agent collaboration, SOP guidance, search, voting, or external causal priors. We intentionally use this basic setting to isolate the intrinsic limitations of monolithic LLM reasoning with tool-augmented evidence collection. The detailed definition of this CoT baseline is provided later in the experimental setup. We conduct a post-mortem analysis on 100 failed cases from the AIOps 2022 dataset, in which the model does not correctly identify the root cause. The failures can be systematically categorized into two dominant types.

\textbf{Evidence Dilution.} This failure mode occurs when the model does encounter root-cause-related signals during its reasoning process, but these signals are progressively weakened or overshadowed by irrelevant or spurious evidence. We observe 43 such cases, which can be further decomposed into two subtypes:

\emph{- Lost Root Cause} (8 cases). The reasoning trajectory contains explicit evidence pointing to the true root cause, yet this evidence is not reflected in the final RCL output. The model ``forgets'' or abandons the correct hypothesis as the reasoning context grows, typically after being distracted by later observations.

\emph{- Demoted Root Cause} (35 cases). The final RCL result does include the correct root cause, but it is ranked below other hypotheses. The model accumulates a large volume of heterogeneous evidence and fails to maintain a clear causal priority, leading to noisy hypothesis aggregation and suboptimal ranking.

These patterns indicate that monolithic LLM reasoning over long, interleaved traces, logs, and metrics lacks a mechanism for evidence isolation and preservation. As the reasoning context expands, salient causal signals are diluted by tangential observations, causing the model to lose focus on the true fault origin.

\textbf{Shallow Reasoning.} In 57 cases, the model terminates its reasoning before reaching the true root cause. Instead of drilling down along the causal chain, it settles on surface-level symptoms (e.g., downstream service failures, error codes, or abnormal metrics) and prematurely outputs a diagnosis. This behavior reflects an inherent tendency of single-agent LLMs to favor early plausible explanations over deeper causal exploration, especially under long-horizon reasoning. Without structural guidance, the model often fails to traverse the full dependency path from symptom to origin, resulting in diagnoses that are descriptively correct but causally incomplete.

To further quantify the impact of shallow reasoning, we conduct a controlled experiment by varying the maximum number of reasoning rounds allowed for the CoT baseline. Specifically, we impose a fixed upper bound on the number of thought-action-observation iterations and force the model to perform exactly this many rounds, regardless of whether it claims to have reached a conclusion earlier. After each round, we record the intermediate and final root cause localization results. In this experiment, the backbone model is Qwen-plus, the dataset is AIOps 2022, and the evaluated method is the above single-agent ReAct-style CoT baseline. We vary the maximum number of reasoning rounds while keeping the prompt format, available tools, input data, and output format unchanged. This setting isolates the effect of reasoning depth and accumulated context from other implementation factors.

\begin{figure}[h]
	\begin{minipage}[b]{1.0\linewidth}
		\centering
		\includegraphics[width=\linewidth]{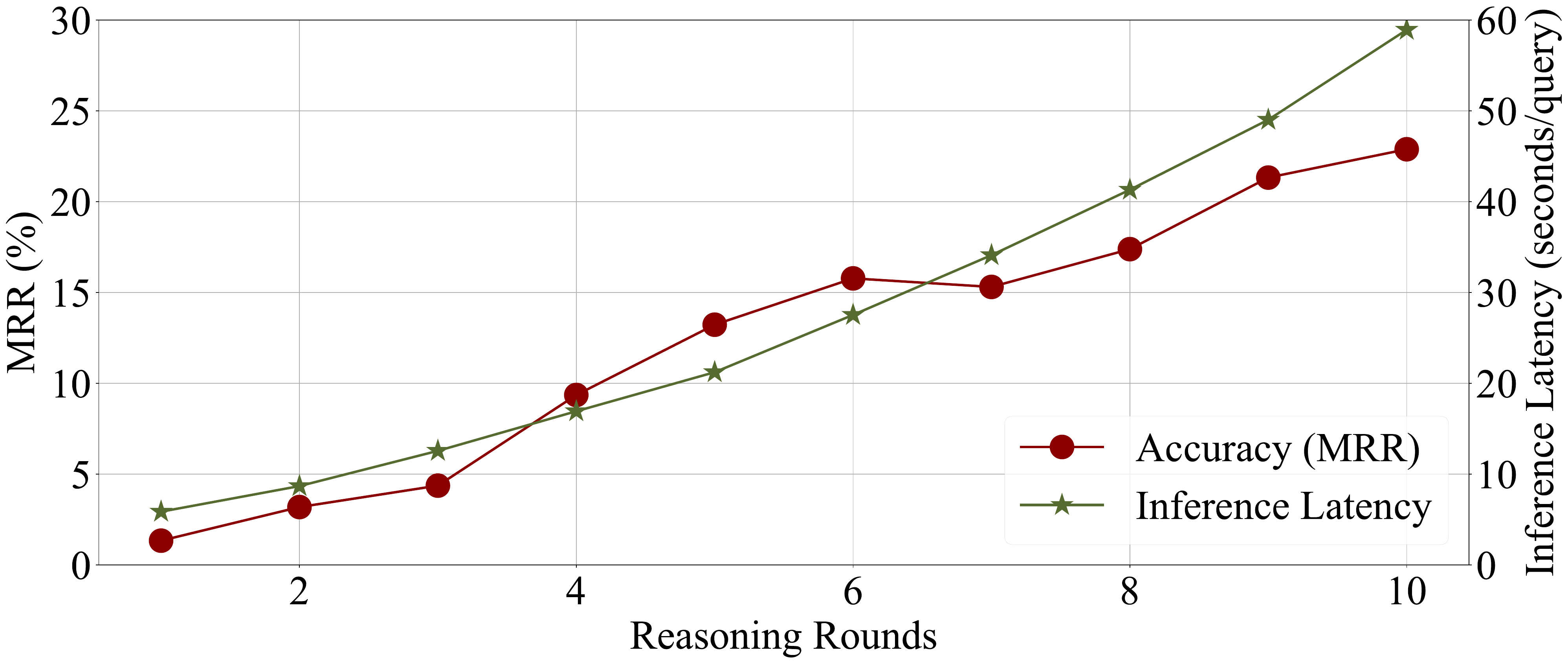}
		\caption{Effect of Reasoning Rounds on RCL Accuracy (MRR) and Latency}
		\label{fig:think-deep-result}
	\end{minipage}
\end{figure}

Figure~\ref{fig:think-deep-result} reports the localization performance of Qwen-plus on the AIOps 2022 dataset under different reasoning rounds. The results exhibit a clear monotonic trend: permitting more rounds of deliberation consistently improves accuracy, confirming that many failures stem from insufficient depth along a serial reasoning path. However, increasing the number of reasoning rounds also increases latency almost linearly. More importantly, because the CoT baseline maintains a single global reasoning trajectory, deeper reasoning also enlarges the context that the model must maintain, compare, and aggregate. This creates a fundamental tension in monolithic LLM-based RCA: deeper exploration can improve accuracy, but it also increases inference cost and exacerbates context explosion, which may dilute critical evidence encountered in earlier reasoning steps.

It is worth noting that prior LLM-based RCA methods have already recognized parts of this problem and proposed useful mitigation strategies. For example, ReAct-style systems such as RCAgent~\cite{wang2024rcagent} use tool invocation to avoid placing all raw data into the initial prompt; SOP-guided systems such as Flow-of-Action~\cite{pei2025flow} constrain the action space using expert procedures; multi-agent systems such as mABC~\cite{zhang2024mabc} introduce collaboration and voting; and hybrid systems such as GALA~\cite{tian2025gala} incorporate statistical or causal priors to guide LLM reasoning. These designs improve controllability and reduce unnecessary exploration to different degrees. However, they do not fundamentally eliminate the accumulated-context problem in long-horizon RCA. In many cases, heterogeneous evidence collected from traces, logs, metrics, and multiple services is still mixed into a centralized reasoning state or aggregated at a coarse granularity. Our contribution is therefore not to claim that context explosion is entirely new, but to reframe it as a central and systematic obstacle for LLM-based root cause localization, and to address it through the RCLAgent design described in the following section.

\textit{Case Study of Context Explosion.}
We present a concrete failure case from the AIOPS 2022 dataset (2022-03-21) to illustrate how evidence dilution manifests under single-trace analysis. The injected fault is \emph{k8s container network packet loss} on pod \texttt{recommendationservice-2}. The anomalous request trace spans 26 nodes across 9 services and pods, with \texttt{frontend-0} as the entry span recording a total latency of 48,276,ms. The root cause signal is structurally clear at the span level: the \texttt{frontend-0 $\rightarrow$ RecommendationService/ListRecommendations} edge accounts for 48,213,ms (99.9\%) of end-to-end latency, while all other branches—seven \texttt{ProductCatalogService} calls (4–10,ms each), \texttt{CartService} (2,ms), \texttt{CurrencyService} (0–2,ms), and \texttt{AdService} (6,ms)—remain entirely normal. Crucially, the diagnostic signal requires \emph{cross-span inference}: whereas the caller-side span records 48,213,ms, the internal span of \texttt{recommendationservice-2} records only \textbf{2,ms}, indicating that the service processes requests correctly but packets are dropped in the network path between \texttt{frontend-0} and \texttt{recommendationservice-2}.

Despite this evidence being present in the trace, the CoT baseline ranks \texttt{productcatalogservice} as the top-1 root cause, with its three pod instances filling positions 1–4, while the true faulty pod \texttt{recommendationservice-2} is demoted to rank~\#7. Because the CoT baseline follows a monolithic reasoning trajectory, it invokes metric and log tools for all nine services encountered in the trace. \texttt{productcatalogservice} appears eight times across three distinct pods, generating three times as many tool call results as \texttt{recommendationservice-2}. As the context accumulates, the critical cross-span latency discrepancy—which can only be identified by jointly comparing the caller-side and callee-side duration of the \emph{same} RPC call—is progressively overshadowed by a large volume of normal but voluminous metric observations from other services. The model, without a structural mechanism to isolate and preserve this span-level causal relationship, converges on the most prominently represented service rather than the true origin of the fault.

This case illustrates the evidence dilution problem observed in our post-mortem analysis. The model does not completely fail to access root-cause evidence; instead, the evidence is gradually weakened in a long, interleaved reasoning context containing many heterogeneous observations. This observation motivates the trace-aligned decomposition and evidence-preserving design of RCLAgent, which we describe in detail in the next section.

\begin{center}
	\begin{tcolorbox}[colback=gray!10,
		colframe=black,
		width=\linewidth,
		arc=1mm, auto outer arc,
		boxrule=0.5pt,
		top=2pt, 
		bottom=2pt, 
		left=2pt,
		right=2pt
		]
		\textbf{Summary.} Existing LLM-based methods for root cause localization primarily suffer from two limitations: \emph{evidence dilution} caused by context explosion, where excessive or accumulated context overwhelms the model and critical causal signals are overlooked; and \emph{shallow reasoning} caused by serial reasoning, where insufficient exploration prevents full traversal of the causal chain, leading to inaccurate root cause identification.
	\end{tcolorbox}
\end{center}

\section{RCLAgent}

\begin{figure*}[htbp]
	\centering
	\includegraphics[width=1\linewidth]{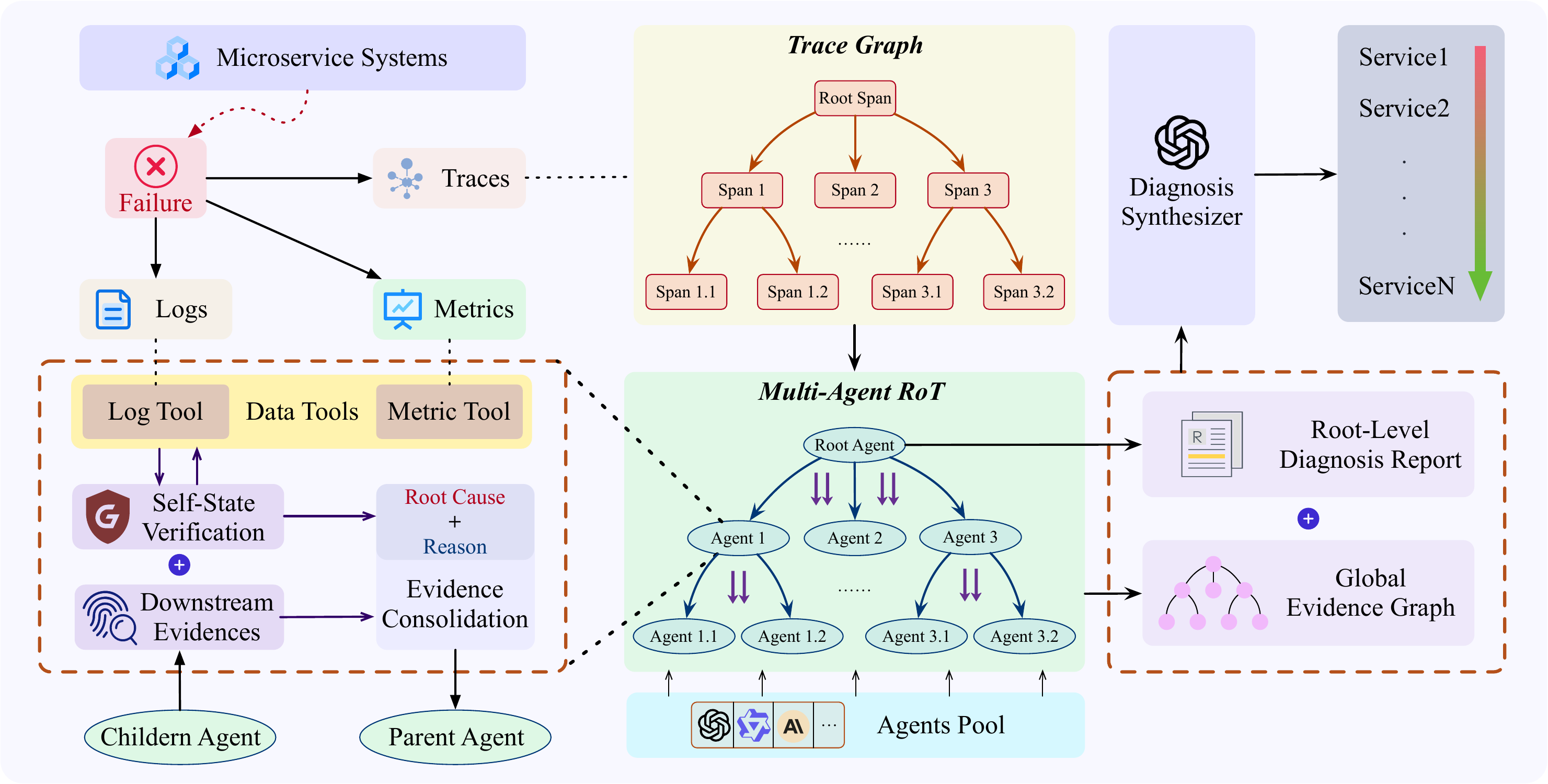}
	\caption{Architecture of RCLAgent}
	\label{fig: architecture}
\end{figure*}

Our empirical study highlights both how human SREs localize the root cause of failures and the limitations of existing LLM-based methods for root cause localization. Motivated by these insights, we introduce \textbf{RCLAgent}, an in-depth framework for root cause localization in microservice systems that implements multi-agent recursion-of-thought with parallel reasoning. Figure~\ref{fig: architecture} illustrates the overall architecture of RCLAgent.

RCLAgent decomposes the diagnostic process according to the structure of the trace graph to implement multi-agent recursion-of-thought (RoT). Specifically, each span is assigned a dedicated agent, and agents are organized recursively following the graph topology. Each agent is instantiated from an \emph{Agents Pool}, which caps the number of concurrently active agents, enabling parallel reasoning across the graph. At the top, the Root Agent generates a root-level diagnosis report. Meanwhile, the entire agent graph produces a \emph{Global Evidence Graph}, in which each agent outputs a compact, structured summary of its local reasoning. Finally, a \emph{Diagnosis Synthesizer} combines the root-level report and the global evidence graph to produce a ranked list of root causes, ordered by likelihood.

The design of RCLAgent directly targets the two failure modes identified in RQ2. First, to mitigate context explosion and evidence dilution, RCLAgent avoids accumulating all trace, log, metric, and reasoning evidence in a single global context. Instead, each Dedicated Agent reasons within a bounded span-level context and emits only compact structured evidence. Evidence Consolidation further prevents raw observations from being propagated upward without abstraction, while the Global Evidence Graph preserves fine-grained local hypotheses so that weak root-cause signals are not lost during recursive aggregation. Second, to mitigate shallow reasoning caused by serial exploration, RCLAgent follows the trace topology and recursively assigns diagnostic responsibilities to downstream agents. Independent branches can be explored in parallel through the Agents Pool, allowing the framework to investigate deeper causal paths without forcing all reasoning steps into one long sequential trajectory.

\subsection{Data Tools}

Data tools serve as the interface between the system and heterogeneous runtime data sources, providing agents with on-demand access to relevant information together with lightweight preprocessing and filtering. In general, a data tool encapsulates a specific data retrieval or analysis capability. In this work, we employ three types of data tools: a Trace Tool, a Metrics Tool, and a Log Tool. The Trace Tool is used exclusively to construct the trace graph and operates independently of agent reasoning. In contrast, the Metrics Tool and Log Tool are invoked by agents during diagnosis.

\subsubsection{Trace Tool} 

Trace data is the most fundamental source for root cause localization, as it records the sequence of calls between various services. However, since each request generates a complete invocation path, the volume of trace data can be enormous, making it difficult for RCLAgent to process such extensive context. To address this, the trace tool is designed to filter and retrieve only the relevant subset of trace data.

\begin{equation}
	T(s) = \left\{ \langle t, s', svc, op, d, \sigma \rangle \; \middle| \; s' \in \mathcal{C}(s) \right\}
	\label{eq: trace-agent}
\end{equation}

Formally, given a span identifier $s$, the trace tool returns a set of child spans along with their associated metadata. We denote this function as Equation~\ref{eq: trace-agent}, where $t$ denotes the timestamp, $s'$ represents the child span identifier, $svc$ is the service name, $op$ is the operation name, $d$ is the duration, $\sigma$ is the status code, and $\mathcal{C}(s)$ is the set of all child spans for the given span $s$.

\subsubsection{Metric Tool}

Metrics reflect the runtime states of system components. In a mature microservice environment, thousands of metrics are continuously collected, often resulting in a data volume that exceeds that of trace data. Our empirical analysis further shows that during anomalous periods, the majority of metrics remain stable and do not exhibit meaningful fluctuations. Directly exposing all metrics to LLM-based agents is therefore inefficient and unnecessary.

To address this issue, we design a dedicated Metric Tool that selectively retrieves only metrics with statistically significant deviations. Specifically, for each metric, its historical mean and standard deviation are estimated over a relatively long reference window to capture stable baseline behavior. Based on this global statistical profile, the Metric Tool then applies an $n$-sigma test within a localized time window centered at the target timestamp to identify abnormal fluctuations.

\begin{equation}
	|m(t) - \mu_m| > n \times \sigma_m
	\label{eq: n-sigma}
\end{equation}

Formally, given a target timestamp $t_0$ and a key component $C$ (e.g., a pod or service), let $\mathcal{M}(C)$ denote the set of metrics associated with $C$ and its related components, including the hosting node in the case of pods. For each metric $m \in \mathcal{M}(C)$, we compute its historical mean $\mu_m$ and standard deviation $\sigma_m$ using a long-term observation window preceding the anomaly. The Metric Tool then examines metric values within the local interval $[t_0 - \delta, t_0 + \delta]$. If there exists any timestamp $t$ in this interval satisfying the $n$-sigma criterion defined in Equation~\ref{eq: n-sigma}, the metric is marked as anomalous.

For all detected anomalous metrics, the Metric Tool returns their fluctuation trajectories within the local window, formally defined as Equation~\ref{eq: metric-agent}.

\begin{equation}
	Q(t_0, \delta, C) = 
	\left\{ 
	m(t) \;\middle|\; 
	\begin{array}{c}
		m \in \mathcal{M}(C),\; t \in [t_0 - \delta, t_0 + \delta] \\
		|m(t) - \mu_m| > n \times \sigma_m
	\end{array} 
	\right\}
	\label{eq: metric-agent}
\end{equation}

Notably, the corresponding $n$-sigma criterion is not used to define failures themselves; rather, it serves only as a lightweight heuristic for filtering anomalous metrics that may provide supporting evidence during diagnosis.

\subsubsection{Log Tool}

Log data provide rich textual evidence of system events and component behaviors, playing a crucial role in root cause localization. However, production systems typically generate logs at a massive scale, while only a small fraction is relevant to a specific incident. Directly exposing raw logs to LLM-based agents is therefore inefficient and may dilute useful evidence.

To address this challenge, we design a dedicated Log Tool that selectively retrieves logs most likely to be indicative of anomalous behaviors for a given component and time window.

Formally, given a target timestamp $t_0$ and a key component $C$ (e.g., a pod or service), let $\mathcal{L}(C)$ denote the set of all log entries associated with $C$ and its related entities. The Log Tool returns a filtered subset of logs within the local interval $[t_0 - \delta, t_0 + \delta]$ that satisfy a predefined relevance criterion, as defined in Equation~\ref{eq: log-agent}.

\begin{equation}
	L(t_0, \delta, C) = 
	\left\{ 
	l \in \mathcal{L}(C) \;\middle|\; 
	t(l) \in [t_0 - \delta, t_0 + \delta] \wedge \phi(l) = 1 
	\right\}
	\label{eq: log-agent}
\end{equation}

Here, $t(l)$ denotes the timestamp of log entry $l$, and $\phi(l)$ is a binary relevance function indicating whether $l$ is potentially informative for anomaly analysis. In practice, $\phi(l)$ is determined using lightweight heuristics, such as log severity levels, error codes, keyword patterns, and correlations with alerts or trace-level anomalies. By filtering logs in this manner, the Log Tool significantly reduces the input context size while preserving the most informative evidence for subsequent recursive reasoning.

\subsection{Multi-Agent RoT}

Multi-Agent RoT decomposes the root cause localization process along the structure of the trace graph. Specifically, each span in the trace graph is assigned a dedicated agent, and agents are organized recursively following the graph topology. Each agent is instantiated from an \emph{Agents Pool}, which limits the number of concurrently active agents and enables efficient parallel reasoning across the graph. At the top of the hierarchy, a Root Agent aggregates the outcomes of subordinate agents and produces a Root-Level Diagnosis Report.

\subsubsection{Dedicated Agent}

Each Dedicated Agent performs \emph{self-state verification} by invoking the Log Tool and the Metric Tool. Except for leaf agents and the Root Agent, it recursively collects downstream evidence from its child agents. The agent then integrates the results of self-state verification with the collected downstream evidence through an \emph{evidence consolidation} process, and forwards the consolidated evidence to its parent agent.

\textbf{Self-State Verification.} This process encourages each Dedicated Agent to reason exclusively over the data associated with its assigned span, aiming to determine whether the span itself
exhibits intrinsic anomalous behavior.

To this end, the agent selectively invokes the Log Tool and the Metric Tool to retrieve
relevant evidence, and then synthesizes its observations into a compact summary. The resulting summary serves two purposes: (1) it is later fused with downstream evidence during the Evidence Consolidation phase, and (2) it constitutes an atomic evidence unit in the construction of the Global Evidence Graph.

Formally, let $s$ denote a span with associated trace attributes $\mathcal{T}(s)$. Let $\mathcal{Q}_{\text{log}}$ and $\mathcal{Q}_{\text{metric}}$ denote the Log Tool and Metric Tool, respectively, which can be invoked by the agent to query auxiliary evidence related to $s$. Self-State Verification is modeled as a tool-augmented reasoning function as Equation~\ref{eq:f-self}, where $e_s$ is a structured self-state evidence summary defined as Equation~\ref{eq:e_s}.

\begin{subequations}
	\begin{equation}
		f_{\text{self}} :
		\big(
		\mathcal{T}(s),
		\mathcal{Q}_{\text{log}},
		\mathcal{Q}_{\text{metric}}
		\big)
		\rightarrow e_s
		\label{eq:f-self}
	\end{equation}
	\begin{equation}
		e_s = \langle s_{\text{id}}, s_{\text{svc}}, a_s, k_s, h_s \rangle
		\label{eq:e_s}
	\end{equation}
\end{subequations}

Here, $s_{\text{id}}$ and $s_{\text{svc}}$ denote the span identifier and service name,
respectively; $a_s \in \{\text{true}, \text{false}\}$ indicates whether the span is deemed abnormal;
$k_s$ summarizes the key observed symptoms; and $h_s$ represents a hypothesis explaining the potential fault or justifying normality.

Figure~\ref{fig:self-state-verification} illustrates the prompt template used to elicit
self-state verification from each agent.

\begin{figure}[htbp]
	\centering
	\begin{tcolorbox}[colback=gray!10, colframe=black, width=\linewidth, arc=1mm, auto outer arc, boxrule=0.5pt, top=2pt, bottom=2pt, left=2pt, right=2pt]
		\textbf{System:} \\
		You are a Root Cause Localization agent in a microservice system.
		A user-reported failure has occurred. Your task is to analyze logs and metrics 
		to identify the DEEPEST ROOT CAUSE SERVICE that initiated the failure chain. \\
		\textbf{User:} \\
		Analyze the following span for root cause: \{$\mathcal{T}(s)$\}, \\
		You have the following data tools to call: \{$\mathcal{Q}_{\text{log}}$, $\mathcal{Q}_{\text{metric}}$\}, \\
		Summarize self-state evidence as JSON:
		\{
			"span\_id": "...",
			"service\_name": "...",
			"is\_abnormal": true/false,
			"key\_symptoms": "brief string",
			"hypothesis": "why it might be faulty or not"
		\}
	\end{tcolorbox}
	\caption{The Prompt for Self-State Verification}
	\label{fig:self-state-verification}
\end{figure}

\textbf{Evidence Consolidation.} Evidence Consolidation integrates the local self-state evidence produced by an agent with the downstream evidences propagated from its child agents, in order to form a localized root cause hypothesis to be passed upward in the agent hierarchy. This process enables recursive aggregation of causal signals while preventing the uncontrolled propagation of raw observations.

Specifically, each Dedicated Agent first collects a set of downstream evidences generated by its child agents, and then jointly reasons over these evidences together with its own self-state summary $e_s$. Based on their relative strength and causal consistency, the agent determines whether the root cause should be attributed to one of its children, to itself, or deferred upward for further analysis by the parent agent.

Formally, let $s$ denote the current span, and let $\mathcal{E}_{\downarrow}(s) = \{ e_{c_1}, \ldots, e_{c_k} \}$ denote the set of evidence summaries propagated from its $k$ child spans. Evidence Consolidation is modeled as a recursive synthesis function as Equation~\ref{eq:f_cons}, where $\hat{e}_s$ is a consolidated local diagnosis summary produced for the parent agent.

\begin{subequations}
	\begin{equation}
		f_{\text{cons}} :
		\big(
		e_s,\;
		\mathcal{E}_{\downarrow}(s)
		\big)
		\rightarrow \hat{e}_s
		\label{eq:f_cons}
	\end{equation}
	\begin{equation}
		\hat{e}_s =
		\langle
		s_{\text{id}},\;
		s_{\text{svc}},\;
		r_s,\;
		r_s^{\text{reason}},\;
		r_s^{\text{conf}}
		\rangle
		\label{eq:hat_e_s}
	\end{equation}
\end{subequations}

Here, $r_s$ denotes the locally inferred root cause candidate (either a child service, the current service itself, or \texttt{self}); $r_s^{\text{reason}}$ provides a concise causal rationale; and $r_s^{\text{conf}} \in [0,1]$ indicates the agent’s confidence in this hypothesis.

\begin{figure}[htbp]
	\centering
	\begin{tcolorbox}[colback=gray!10, colframe=black, width=\linewidth, arc=1mm, auto outer arc, boxrule=0.5pt, top=2pt, bottom=2pt, left=2pt, right=2pt]
		\textbf{User:} \\
		Downstream evidences from children: \{$\mathcal{E}_{\downarrow}(s)$\} \\
		Your own self-evidence: \{$e_s$\} \\
		Task: Synthesize these evidences into a locally consolidated root cause hypothesis
		to be propagated to your parent agent. \\
		Output format (JSON only):
		\{
		"span\_id": "...",
		"service\_name": "...",
		"local\_root\_cause": "service name or 'self'",
		"reason": "...",
		"confidence": 0.0-1.0
		\}
	\end{tcolorbox}
	\caption{The Prompt for Evidence Consoliadation}
	\label{fig:evidence-consoliadation}
\end{figure}

Intuitively, if a child agent exhibits strong and consistent evidence of being the root cause, the agent propagates that hypothesis upward. Otherwise, if the current span itself is deemed abnormal, the agent proposes itself as the local root cause. Figure~\ref{fig:evidence-consoliadation} illustrates the prompt template used to elicit evidence consolidation from each agent.

\subsubsection{Agent Pools}

To support scalable parallel reasoning while avoiding uncontrolled resource consumption, RCLAgent introduces an \emph{Agents Pool} mechanism to regulate the instantiation and execution of Dedicated Agents. Rather than spawning an unbounded number of agents along the trace graph, all agents are created and scheduled through a centralized pool with a fixed capacity.

\begin{equation}
	|\mathcal{A}_{\text{active}}| \leq K
	\label{eq:agent_pool}
\end{equation}

Formally, let $\mathcal{P}$ denote an Agents Pool with a maximum capacity $K$. At any time, the number of concurrently active agents satisfies Equation~\ref{eq:agent_pool}, where $\mathcal{A}_{\text{active}}$ is the set of agents currently executing. When the pool reaches capacity, newly requested agents are queued and instantiated only after active agents complete and release their slots.

\subsection{Diagnosis Synthesizer}

Built on top of the Multi-Agent RoT process, the Diagnosis Synthesizer is responsible
for producing the final ranked list of root cause candidates. Rather than relying on a single source of evidence, it jointly reasons over two complementary representations: (1) a \emph{Root-Level Diagnosis Report} that summarizes recursive hypothesis propagation, and (2) a \emph{Global Evidence Graph} that preserves fine-grained self-state evidences across all spans.

\subsubsection{Root-Level Diagnosis Report}

The Root-Level Diagnosis Report is the final diagnostic outcome produced by the Root Agent after recursively aggregating evidences propagated along the agent graph. From a formal perspective, this report can be modeled as a root-level evidence tuple, which follows the same unified evidence representation used by all agents, but is instantiated at the root of the reasoning hierarchy.

Specifically, the Root-Level Diagnosis Report is defined as $\hat{e}_{\text{root}} = \langle s_{\text{id}},\; s_{\text{svc}},\; r_{\text{root}},\; r_{\text{root}}^{\text{reason}},\; r_{\text{root}}^{\text{conf}} \rangle $, where this tuple is a special case of the general evidence formulation in Equation~\ref{eq:hat_e_s}, corresponding to the Root Agent.

Here, $s_{\text{id}}$ and $s_{\text{svc}}$ identify the root span (or service) associated with the anomalous execution trace. The term $r_{\text{root}}$ denotes the synthesized root cause hypothesis, obtained through recursive Evidence Consolidation over all descendant agents. $r_{\text{root}}^{\text{reason}}$ provides a structured diagnostic rationale, which abstracts and summarizes the multi-level reasoning process, while $r_{\text{root}}^{\text{conf}}$ quantifies the Root Agent’s confidence in the proposed diagnosis.

\subsubsection{Global Evidence Graph}

While the Root-Level Diagnosis Report provides a compact diagnostic summary, it inevitably abstracts away fine-grained observational details. To preserve the full evidential context, RCLAgent constructs a Global Evidence Graph that aggregates self-state verification outputs from all Dedicated Agents across the trace graph.

Formally, the Global Evidence Graph is defined as a directed attributed graph shown in Equation~\ref{eq:GEG}, where each node $v \in V$ corresponds to a span $s$ in the trace graph, each edge $(v_i, v_j) \in E$ follows the parent--child dependency relations in the trace graph, and $\Phi$ is a node attribution function that maps each node to its self-state evidence summary $\Phi(v_s) = e_s$.

\begin{equation}
	\mathcal{G}_E = (V, E, \Phi)
	\label{eq:GEG}
\end{equation}

Here, $e_s$ is the structured self-state evidence produced by Self-State Verification
(Equation~\ref{eq:e_s}), where each node explicitly stores the abnormality judgment, key symptoms, and local causal hypothesis associated with the corresponding span. Structurally, $\mathcal{G}_E$ is isomorphic to the trace graph, but semantically it represents a distributed evidential field rather than an execution topology. Each node encodes localized diagnostic knowledge, and edges preserve causal and temporal dependencies between evidences.

Unlike the Root-Level Diagnosis Report, which reflects recursively propagated conclusions, the Global Evidence Graph serves as a bottom-up evidential substrate. It enables the Diagnosis Synthesizer to: (i) cross-validate root-level hypotheses against distributed observations, (ii) detect inconsistencies between local and global inferences, and (iii) recover weak or diluted causal signals that may be suppressed during hierarchical aggregation.

Ultimately, the Diagnosis Synthesizer jointly reasons over the Root-Level Diagnosis Report and the Global Evidence Graph, yielding the final diagnostic decision.

\begin{equation}
	\mathcal{F}_{\text{synth}} :
	\big( \hat{e}_{\text{root}}, \mathcal{G}_E \big)
	\rightarrow \mathcal{R}
	\label{eq:synthesizer}
\end{equation}

Formally, it is modeled as a mapping shown in Equation~\ref{eq:synthesizer}, where $\mathcal{R}$ denotes the final ranked list of root cause candidates.

\section{Evaluation}

To evaluate RCLAgent, we conduct a series of experimental studies to investigate the following research questions:

\begin{itemize} 
	\item \textbf{EV-RQ1:} How does RCLAgent perform in root cause localization accuracy compared to baseline LLMs and state-of-the-art failure localization approaches?
	\item \textbf{EV-RQ2:} How efficient is RCLAgent in root cause localization when compared with LLM-based baseline approaches?
	\item \textbf{EV-RQ3:} How does the choice of different LLM backbones affect the root cause localization performance of RCLAgent?
	\item \textbf{EV-RQ4:} What is the contribution of each component of RCLAgent to its overall accuracy?
	\item \textbf{EV-RQ5:} How sensitive is RCLAgent's performance to key hyperparameters?
\end{itemize}

\subsection{Experimental Setup}

\subsubsection{Dataset}

We evaluate RCLAgent on three representative benchmarks: AIOPS 2022~\cite{aiops2022championship}, Augmented-TrainTicket~\cite{yu2023nezha}, and RCAEval~\cite{pham2025rcaeval}.

\textbf{AIOps 2022} benchmark is a large-scale, real-world dataset collected from a production-grade, microservices-based e-commerce system, reflecting realistic operational complexities and failure patterns. It has been used in several notable works, such as Grace~\cite{ren2023grace} and TVDiag~\cite{xie2025tvdiag}. The monitored system consists of 7 microservices deployed across 44 pods and 6 nodes, with traces ranging from hundreds to thousands of spans. The dataset includes four business metrics, 400 performance metrics, as well as traces and logs.

\textbf{Augmented-TrainTicket} is a microservices-based train ticket booking system comprising 41 microservices. Built upon the open-sourced TrainTicket system, Nezha~\cite{yu2023nezha} augments it with comprehensive observability, including traces, metrics, and logs, enabling fine-grained root cause analysis.

\textbf{RCAEval} is an open-source benchmark that provides both datasets and a unified evaluation environment for root cause analysis in microservice systems. It contains three comprehensive datasets with a total of 735 failure cases, collected from three distinct microservice systems. Since RCLAgent requires the joint availability of traces, metrics, and logs, we adopt the RE2-OB subdataset in our experiments.

\subsubsection{Baseline Approaches}

We compared RCLAgent with the following eight root cause localization approaches, which can be categorized into two groups: non-LLM-based, and LLM-based methods.

\textbf{non-LLM-based:} Some approaches rely primarily on trace data. CRISP~\cite{zhang2022crisp} represents traces as critical paths and applies lightweight heuristics to identify root-cause instances. TraceContrast~\cite{zhang2024trace} leverages sequence representations, contrastive sequential pattern mining, and spectrum analysis to localize multi-dimensional root causes. TraceRank~\cite{yu2023tracerank} combines spectrum analysis with a PageRank-based random walk algorithm to pinpoint anomalous services. MicroRank~\cite{yu2021microrank} constructs a trace coverage tree to model dependencies between requests and service instances, and applies PageRank to score candidate root causes. Other methods focus on metrics data. RUN~\cite{lin2024root} employs time-series forecasting for neural Granger causal discovery and integrates a personalized PageRank algorithm to efficiently recommend the top-$k$ root causes. Microscope~\cite{lin2018microscope} constructs causal graphs and applies a depth-first search strategy to identify front-end anomalies. CausalRCA~\cite{xin2023causalrca} adopts a gradient-based causal structure learning approach to generate weighted causal graphs, followed by a root-cause inference procedure to localize anomalous metrics. Some approaches leverage multimodal data. Eadro~\cite{lee2023eadro} jointly models intra-service behaviors and inter-service dependencies using traces, logs, and metrics, and exploits shared knowledge across these two modeling phases through multi-task learning.

\textbf{LLM-based:} mABC~\cite{zhang2024mabc} adopts a multi-agent, blockchain-inspired collaboration framework, in which multiple LLM-based agents follow a structured workflow and coordinate through blockchain-style voting mechanisms. RCAgent~\cite{wang2024rcagent} employs a ReAct-based reasoning-and-acting paradigm to jointly process multiple alerts, where a controller agent iteratively executes a thought–action–observation loop and invokes external tools for collective analysis. GALA~\cite{tian2025gala} proposes a multimodal root cause localization framework that integrates statistical causal inference with LLM-driven iterative reasoning, enabling joint reasoning over heterogeneous operational data. CoT-based approaches~\cite{wei2022chain} apply a naive chain-of-thought reasoning strategy, typically handling alerts in isolation and generating conclusions in a sequential manner.

\begin{table*}[tbp]
	\setlength{\tabcolsep}{4.0pt}
	\centering
	\caption{Root Cause Localization Accuracy Compared with SOTA Methods (R$k$ denotes Recall@$k$).}
	\label{tab: accuracy}
	\begin{tabular}{c|c|ccccc|ccccc|ccccc}
		\toprule
		\multirow{2}{*}{\textbf{Paradigm}} & \multirow{2}{*}{\textbf{Method}} & \multicolumn{5}{c|}{AIOPS 2022} & \multicolumn{5}{c|}{TrainTicket} & \multicolumn{5}{c}{RCAEval} \\
		\\[-2.0ex]
		\cline{3-17}
		\\[-1.5ex]
		~ & ~ & \textit{\textbf{R1}} & \textit{\textbf{R3}} & \textit{\textbf{R5}} & \textit{\textbf{R10}} & \textit{\textbf{MRR}} & \textit{\textbf{R1}} & \textit{\textbf{R3}} & \textit{\textbf{R5}} & \textit{\textbf{R10}} & \textit{\textbf{MRR}} & \textit{\textbf{R1}} & \textit{\textbf{R3}} & \textit{\textbf{R5}} & \textit{\textbf{R10}} & \textit{\textbf{MRR}} \\
		\midrule
		\multirow{9}{*}{\textit{non-LLM-based}} & CRISP & 3.92 & 29.79 & 48.88 & 51.01 & 18.68 & 31.45 & 41.77 & 48.19 & 55.26 & 38.53 & 14.31 & 25.99 & 39.13 & 47.65 & 24.83 \\
		~ & TraceConstract & 24.06 & 36.51 & 37.69 & 40.81 & 30.29 & 48.37 & 55.33 & 63.78 & 78.21 & 54.35 & 39.45 & 45.33 & 56.72 & 60.08 & 45.79 \\
		~ & TraceRCA & 36.43 & 43.13 & 44.56 & 51.88 & 42.19 & 42.73 & 46.98 & 47.17 & 49.32 & 45.34 & 23.34 & 45.51 & 57.39 & 61.38 & 38.17 \\
		~ & MicroRank & 8.45 & 24.57 & 37.24 & 42.89 & 18.42 & 24.15 & 42.10 & 50.16 & 62.33 & 34.73 & 16.23 & 29.39 & 32.58 & 39.13 & 24.73 \\
		~ & RUN & 4.69 & 11.23 & 11.88 & 25.81 & 10.44 & 40.03 & 47.19 & 51.31 & 56.77 & 48.38 & 12.34 & 25.76 & 37.64 & 45.95 & 23.43 \\
		~ & MicroScope & 9.23 & 28.81 & 33.37 & 51.25 & 20.39 & 32.37 & 40.18 & 43.26 & 48.70 & 36.88 & 13.38 & 28.79 & 40.05 & 53.34 & 26.16 \\
		~ & CausalRCA & 12.33 & 29.02 & 35.17 & 49.31 & 24.53 & 38.86 & 49.33 & 53.65 & 56.01 & 45.69 & 23.39 & 44.32 & 58.67 & 66.15 & 36.18 \\
		~ & Eadro & 15.67 & 31.03 & 34.79 & 45.19 & 25.98 & 43.18 & 49.97 & 55.63 & 58.19 & 48.39 & 19.13 & 43.95 & 63.02 & 69.37 & 37.31 \\
		~ & Nezha & 6.08 & 12.93 & 14.07 & 14.07 & 9.42 & \textbf{86.67} & \textbf{97.78} & \textbf{97.78} & \textbf{97.78} & \textbf{90.37} & 00.00 & 00.00 & 00.00 & 1.10 & 0.56 \\
		~ & BARO & 23.96 & 44.11 &50.95 & 58.55 & 36.07 & 0.00 & 5.88 & 23.53 & 100.00 & 17.00 & 18.89 & \textbf{92.22} & \textbf{100.00} & \underline{100.00} & 55.11 \\
		\midrule
		\multirow{4}{*}{\textit{LLM-based}} & CoT & 8.96 & 15.89 & 34.72 & 48.33 & 17.56 & 22.12 & 39.75 & 48.66 & 61.19 & 32.14 & 13.67 & 20.18 & 23.16 & 28.81 & 18.34 \\
		~ & RCAgent & 13.15 & 23.11 & 38.56 & 54.87 & 22.53 & 38.51 & 53.14 & 60.47 & 69.98 & 47.39 & 25.32 & 31.09 & 39.19 & 45.83 & 31.28 \\
		~ & mABC & \underline{51.68} & \underline{60.33} & \underline{65.33} & \underline{78.46} & \underline{58.05} & 57.53 & 74.67 & 83.15 & 90.02 & 67.26 & 39.98 & 42.13 & 44.95 & 47.87 & 41.76 \\
		~ & GALA & 33.88 & 45.75 & 53.64 & 59.73 & 42.06 & 59.66 & 66.78 & 69.34 & 74.83 & 64.28 & \underline{45.59} & \textbf{66.83} & \underline{75.32} & \underline{78.94} & \underline{58.93} \\
		\midrule
		\multicolumn{2}{c|}{RCLAgent (\textit{Claude-3.5-Sonnet})} & 61.39 & 69.58 & 79.39 & 90.10 & 69.73 & 66.36 & 75.42 & 81.13 & 91.18 & 74.08 & 52.31 & 63.15 & 79.39 & 85.96 & 62.97 \\
		\multicolumn{2}{c|}{RCLAgent (\textit{Qwen-3.6-Plus})} & \textbf{65.15} & \textbf{78.79} & \textbf{86.36} & \textbf{95.45} & \textbf{73.24} & \underline{82.35} & \underline{88.24} & \underline{94.12} & \underline{94.12} & \underline{86.47} & \textbf{56.67} & 80.00 & 86.67 & \textbf{100.00} & \textbf{71.03} \\
		\bottomrule
	\end{tabular}
\end{table*}

\subsubsection{Evaluation Metrics}

We use the top-k recall (Recall@k) and mean reciprocal rank (MRR) to evaluate the accuracy of root cause localization following existing works~\cite{zhang2024trace}.

\begin{itemize}
	\item \textbf{Recall@k:} Measures the likelihood that the true root cause appears within the top-k results in the ranked list. Specifically, it indicates whether the root cause is found within the first $k$ predictions. In this paper, we evaluate Recall@1, Recall@5, and Recall@10.
	\item \textbf{MRR:} is the multiplicative inverse of the rank of the root cause in the result list. If the root cause is not included in the top-10 result list, the rank can be regarded as positive infinity. Given a set of fault instances $A$, $Rank_i$ is the $i$ rank of the root cause in the returned list of the $i$th fault instance, MRR is calculated by Equation~\ref{eq: mrr}.
\end{itemize}

\begin{equation}
	MRR = \frac{1}{|A|} \sum_{i=1}^{|A|} \frac{1}{Rank_i}
	\label{eq: mrr}
\end{equation}

Since many existing methods are limited to localizing the root cause only at the pod level, we assume that if these methods correctly identify the pod, the corresponding service to which the pod belongs is considered the root cause, and we treat such predictions as correct.

\subsubsection{Implementation and Settings}

We implement RCLAgent in Python 3.10. Unless otherwise specified, Claude-3.5 Sonnet is used as the underlying LLM engine. For the Metric Tool, the $n$-sigma threshold is set to $n = 3$, the retrieval interval for fluctuating metrics is set to $\delta = 60\text{s}$, and the size of the Agents Pool is fixed at $K = 100$. All experiments are conducted on a Linux server with an x86-64 architecture, equipped with 24 CPU cores running at 2.90 GHz, 400 GB RAM.

\subsection{Accuracy}

We first compare RCLAgent with both non-LLM-based and LLM-based methods in terms of localization accuracy, addressing RQ1. As shown in Table~\ref{tab: accuracy}, we evaluate RCLAgent with two LLM backbones, Claude-3.5-Sonnet and Qwen-3.6-Plus. The results show that RCLAgent outperforms almost all state-of-the-art baselines in terms of MRR, with the only exception being Nezha on the Augmented-TrainTicket dataset.

On the AIOPS 2022 dataset, RCLAgent achieves an MRR of 69.73\%, exceeding the strongest LLM-based baseline, mABC, by 11.68\%, and the best non-LLM-based method, TraceRCA, by 27.54\%. On the Augmented-TrainTicket dataset, RCLAgent attains an MRR of 74.08\%, outperforming mABC by 6.82\% and TraceConstruct by 19.73\%. Notably, RCLAgent performs slightly worse than mABC only on Recall@5, with a marginal gap of 2.02\%. On the RCAEval dataset, RCLAgent reaches an MRR of 62.97\%, surpassing the LLM-based method GALA by 4.04\% and TraceConstruct by 17.18\%. A minor performance drop is observed only on Recall@3, where RCLAgent trails GALA by 3.68\%.

To further verify the robustness of our results and address concerns regarding the margin of improvement, we conduct a paired Wilcoxon signed-rank test comparing RCLAgent with the strongest baselines, including mABC and GALA. The resulting $p$-values for Recall@k ($k=1, 5, 10$) and MRR across all datasets are consistently below $0.05$; specifically, the $p$-value for MRR is $0.012$. These results confirm that the performance gains of RCLAgent are statistically significant.

Although RCLAgent is slightly outperformed by Nezha on Augmented-TrainTicket, Nezha exhibits much weaker performance on AIOPS 2022 and RCAEval, indicating limited cross-dataset generalization. This is mainly because Nezha depends heavily on specific trace-structure changes and modality correlations, which may not hold across different fault types or observability settings. In contrast, RCLAgent integrates heterogeneous evidence through LLM-based reasoning and is less tied to a single dataset-specific structural assumption. We provide a more detailed discussion of this trade-off in Section~\ref{sec:why-llm}.

We also observe that RCLAgent with Qwen-3.6-Plus generally achieves better performance than RCLAgent with Claude-3.5-Sonnet. This suggests that RCLAgent can naturally benefit from stronger LLM backbones. As LLM capabilities continue to improve, the diagnostic accuracy and reasoning quality of RCLAgent are expected to further improve without requiring major changes to the overall framework.

\subsection{Efficiency}

We further evaluate the localization efficiency of RCLAgent in comparison to LLM-based methods, addressing RQ2. Table~\ref{tab: efficiency} reports the average inference time per query across the three datasets.

\begin{table}[htb]
	\setlength{\tabcolsep}{3.0pt}
	\centering
	\caption{Inference Speed Comparison (seconds/query)}
	\label{tab: efficiency}
	\begin{tabular}{c|c|ccc}
		\toprule
		Paradigm & Method & AIOPS 2022 & TrainTicket & RCAEval \\
		\midrule
		\multirow{9}{*}{\textit{non-LLM-based}} & TraceContrast & 2.51 & 2.84 & 2.37 \\
		~ & TraceRCA & \underline{0.93} & \underline{1.02} & \underline{0.87} \\
		~ & MicroRank & 1.47 & 1.63 & 1.38 \\
		~ & RUN & 9.23 & 8.71 & 9.58 \\
		~ & MicroScope & 1.82 & 1.69 & 1.77 \\
		~ & CausalRCA & 23.15 & 24.81 & 22.43 \\
		~ & Eadro & 5.37 & 5.92 & 5.14 \\
		~ & Nezha & 4.93 & 4.81 & 4.87 \\
		~ & BARO & \textbf{0.46} & \textbf{0.71} & \textbf{0.13} \\
		\midrule
		\multirow{5}{*}{\textit{LLM-based}} & CoT & \underline{87.63} & \underline{89.73} & \underline{85.77} \\
		~ & RCAgent & 103.36 & 119.98 & 125.33 \\
		~ & mABC & 173.12 & 195.34 & 158.70 \\
		~ & GALA & 181.19 & 193.33 & 176.37 \\
		~ & RCLAgent (\textit{ours}) & \textbf{41.43} & \textbf{59.55} & \textbf{49.79} \\
		\bottomrule
	\end{tabular}
\end{table}

As shown in Table~\ref{tab: efficiency}, non-LLM-based methods are generally much faster than LLM-based methods in terms of per-query inference time. This is expected because most non-LLM baselines rely on lightweight statistical analysis, graph traversal, or pre-trained scoring models, whereas LLM-based methods require multiple model calls and natural-language reasoning steps. Among the non-LLM baselines, BARO achieves the shortest inference time across all three datasets, followed by TraceRCA. Compared with these lightweight methods, RCLAgent introduces higher runtime latency due to LLM inference and tool-augmented reasoning.

Nevertheless, within the LLM-based category, RCLAgent achieves the best efficiency on all datasets. Compared with CoT, RCAgent, mABC, and GALA, RCLAgent consistently reduces the average inference time by a large margin. Specifically, compared with the strongest LLM-based baseline in terms of speed, namely CoT, RCLAgent reduces the average inference time from 87.63s to 41.43s on AIOPS 2022, from 89.73s to 59.55s on TrainTicket, and from 85.77s to 49.79s on RCAEval. This corresponds to a speedup of approximately 1.49$\times$--2.11$\times$. Compared with more complex agent-based baselines such as mABC and GALA, the reduction is even more substantial.

The efficiency gain of RCLAgent primarily stems from its parallel multi-agent reasoning strategy. By decomposing the trace graph into span-level Dedicated Agents and executing independent branches concurrently under a sufficiently large Agents Pool, RCLAgent avoids the long serial reasoning chains commonly observed in ReAct- and CoT-style frameworks. At the same time, RCLAgent does not collapse the entire diagnostic process into a single inference step. Its workflow remains hierarchically structured along the trace graph: each agent performs localized analysis, propagates compact evidence upward, and the final Diagnosis Synthesizer aggregates the Root-Level Diagnosis Report and Global Evidence Graph. Therefore, although RCLAgent remains slower than lightweight non-LLM methods, it substantially improves the efficiency of LLM-based root cause localization while preserving structured and interpretable reasoning.

Overall, the expanded comparison highlights a practical trade-off. Non-LLM-based methods offer lower per-query latency, but they often depend on dataset-specific assumptions, fixed feature designs, or extensive environment-specific preprocessing. In contrast, RCLAgent incurs higher inference cost but provides stronger generalization ability, flexible evidence integration, and an explicit reasoning process that can assist human SREs in validating and resolving incidents.

\subsection{LLM Backbone Impact}

We next investigate how the choice of LLM backbone influences the root cause localization performance of RCLAgent, addressing RQ3. Experiments are conducted on the AIOPS 2022 dataset, which is further partitioned into six subsets according to cloudbed and date, denoted as $\mathbf{A}$, $\mathbf{B}$, $\mathbf{\Gamma}$, $\mathbf{\Delta}$, $\mathbf{E}$, and $\mathbf{Z}$. Beyond the default Claude-3.5-Sonnet backbone, we evaluate several representative large language models with diverse architectures and training paradigms, including DeepSeek-R1-Qwen (a 32B Qwen2.5 model fine-tuned on distilled reasoning data), Qwen-2.5-Max and Qwen-2.5-Plus (two closed-source API-based models), Llama-3.1-70B, and GPT-4. In addition, we report HumanEval~\cite{chen2021evaluating} scores as a representative reasoning-oriented benchmark to contextualize the general reasoning capabilities of each backbone.

\begin{table}[htb]
	\setlength{\tabcolsep}{2.4pt}
	\centering
	\caption{RCLAgent with different LLMs (MRR)}
	\label{tab: llm-impact}
	\begin{tabular}{c|cccccc|c}
		\toprule
		Model & $\mathbf{A}$ & $\mathbf{B}$ & $\mathbf{\Gamma}$ & $\mathbf{\Delta}$ & $\mathbf{E}$ & $\mathbf{Z}$ & HumanEval \\
		\midrule
		Claude-3.5-sonnet & \textbf{69.03} & \textbf{80.38} & \textbf{88.67} & \underline{61.03} & \textbf{64.39} & \underline{54.88} & \textbf{93.7} \\
		DeepSeek-R1-qwen & 36.13 & 62.10 & \underline{63.89} & \textbf{65.13} & \underline{47.01} & \textbf{59.33} & 89.0 \\
		Qwen-2.5-max & 18.52 & 27.46 & 38.73 & 19.34 & 37.31 & 20.09 & 86.6 \\
		Qwen-2.5-plus & \underline{43.19} & \underline{67.17} & 55.68 & 47.39 & 39.33 & 55.70 & \underline{92.7} \\
		Llama-3.1-70B & 8.31 & 13.68 & 19.42 & 18.03 & 7.09 & 9.13 & 80.5 \\
		GPT-4 & 6.35 & 9.24 & 14.71 & 13.52 & 4.18 & 5.63 & 87.1 \\
		\bottomrule
	\end{tabular}
\end{table}

Table~\ref{tab: llm-impact} summarizes the results. Overall, RCLAgent’s performance exhibits a strong dependence on the capability and task alignment of the underlying LLM. Claude-3.5-Sonnet consistently achieves the best or near-best performance across most subsets, attaining the highest average MRR and outperforming the second-best backbone by 14.13\% on average. This result highlights the importance of robust multi-step reasoning and semantic understanding in enabling effective multi-agent root cause localization.

Notably, DeepSeek-R1-Qwen demonstrates competitive performance on certain subsets. On $\mathbf{\Delta}$ and $\mathbf{Z}$, it achieves MRR scores of 65.13 and 59.33, respectively, slightly surpassing Claude-3.5-Sonnet (61.03 and 54.88) by 4.10\% and 4.45\%. However, this advantage does not generalize to more complex subsets. On $\mathbf{A}$ and $\mathbf{B}$, DeepSeek-R1-Qwen yields substantially lower MRR scores (36.13 and 62.10) than Qwen-2.5-Plus (43.19 and 67.17), which we attribute to the relatively smaller model scale (32B) and its limited capacity for sustained reasoning over long and heterogeneous system contexts.

Interestingly, models with strong performance on general reasoning benchmarks do not necessarily translate to strong performance in RCLAgent. For instance, GPT-4 attains a relatively high HumanEval score (87.1) but performs poorly across all AIOPS subsets, with MRR values consistently below 15. This discrepancy suggests that general-purpose code reasoning ability alone is insufficient; effective integration within RCLAgent additionally requires task-specific alignment, stable tool interaction, and robustness to noisy and structured operational data. Similar trends can be observed for Llama-3.1-70B, which, despite its large parameter size, underperforms across all subsets.

Taken together, these results indicate that while stronger general reasoning capability is a necessary condition for high RCLAgent performance, it is not sufficient. The effectiveness of RCLAgent depends not only on the backbone’s raw reasoning strength, but also on its alignment with tool-augmented, multi-agent workflows and its ability to reason over complex, real-world system telemetry.

These findings also have an important practical implication for deployment. Backbone selection for RCLAgent cannot rely solely on standard general-purpose LLM benchmarks, since different models may exhibit different strengths across datasets, failure types, and telemetry characteristics. In practice, an appropriate model should be selected not only based on general reasoning ability, but also on task-specific factors such as tool-use stability, consistency in multi-step reasoning, robustness to noisy and structured operational data, and the latency-cost trade-off in the target environment. We therefore recommend selecting the backbone through validation on a small set of representative historical incidents from the target system, using end-to-end RCL performance together with operational indicators such as output stability, tool-call reliability, latency, and monetary cost.

\subsection{Ablation Study}

We further examine the contribution of each major component of RCLAgent to its overall localization accuracy, addressing RQ4. Specifically, we evaluate variant models with and without the Root-Level Diagnosis Report (RLDR) and the Global Evidence Graph (GEG).

\begin{figure}[htbp]
	\centering
	\includegraphics[width=1\linewidth]{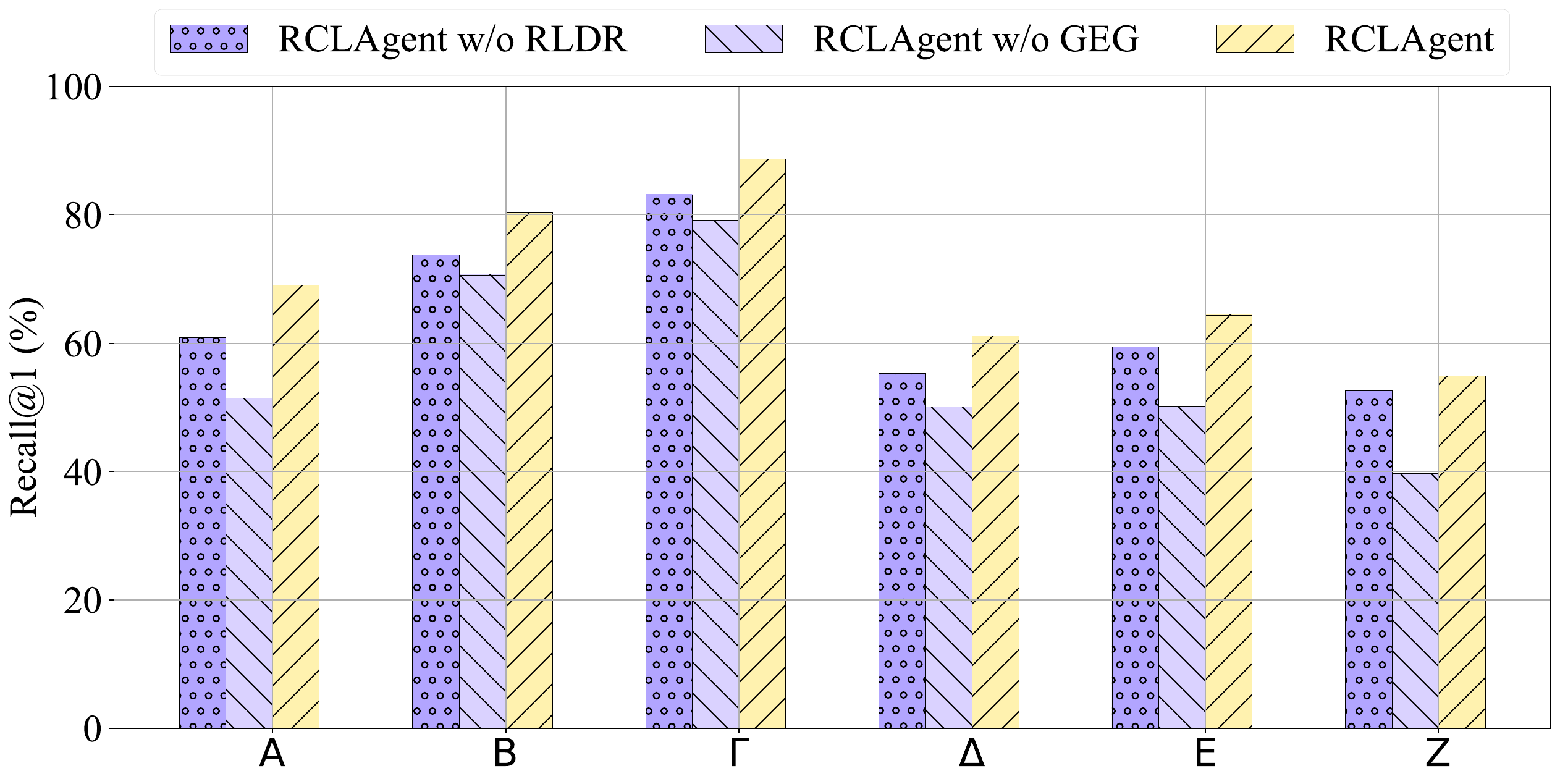}
	\caption{Ablation Experiment}
	\label{fig: ablation}
\end{figure}

As shown in Figure~\ref{fig: ablation}, both components contribute positively to RCLAgent’s root cause localization performance. Among them, RLDR provides a larger individual performance gain: the variant using only RLDR consistently outperforms the one using only GEG, with an average improvement of 7.29\%. This result highlights the critical role of hierarchical evidence propagation and consolidation in accurate diagnosis.

Nevertheless, combining GEG with RLDR yields further performance improvements. Compared to the RLDR-only variant, the full model achieves an additional average gain of 5.54\%. This indicates that GEG effectively complements RLDR by preserving fine-grained, bottom-up evidence that mitigates information loss introduced during recursive aggregation. Together, these results demonstrate that RLDR and GEG are synergistic rather than redundant, jointly enabling more robust and accurate root cause localization.

\subsection{Hyperparameter Analysis}

The parameters in our approach mainly reside in the Data Tools, particularly the Metric Tool. We evaluate the impact of the retrieval interval for fluctuating metrics by varying $\delta$ from $10\text{s}$ to $120\text{s}$. For each configuration, we repeat the experiments with different random seeds and report the mean performance. The variability across runs is visualized using shaded regions in the line plots, indicating one standard deviation.

\begin{figure}[htbp]
	\centering
	\includegraphics[width=1\linewidth]{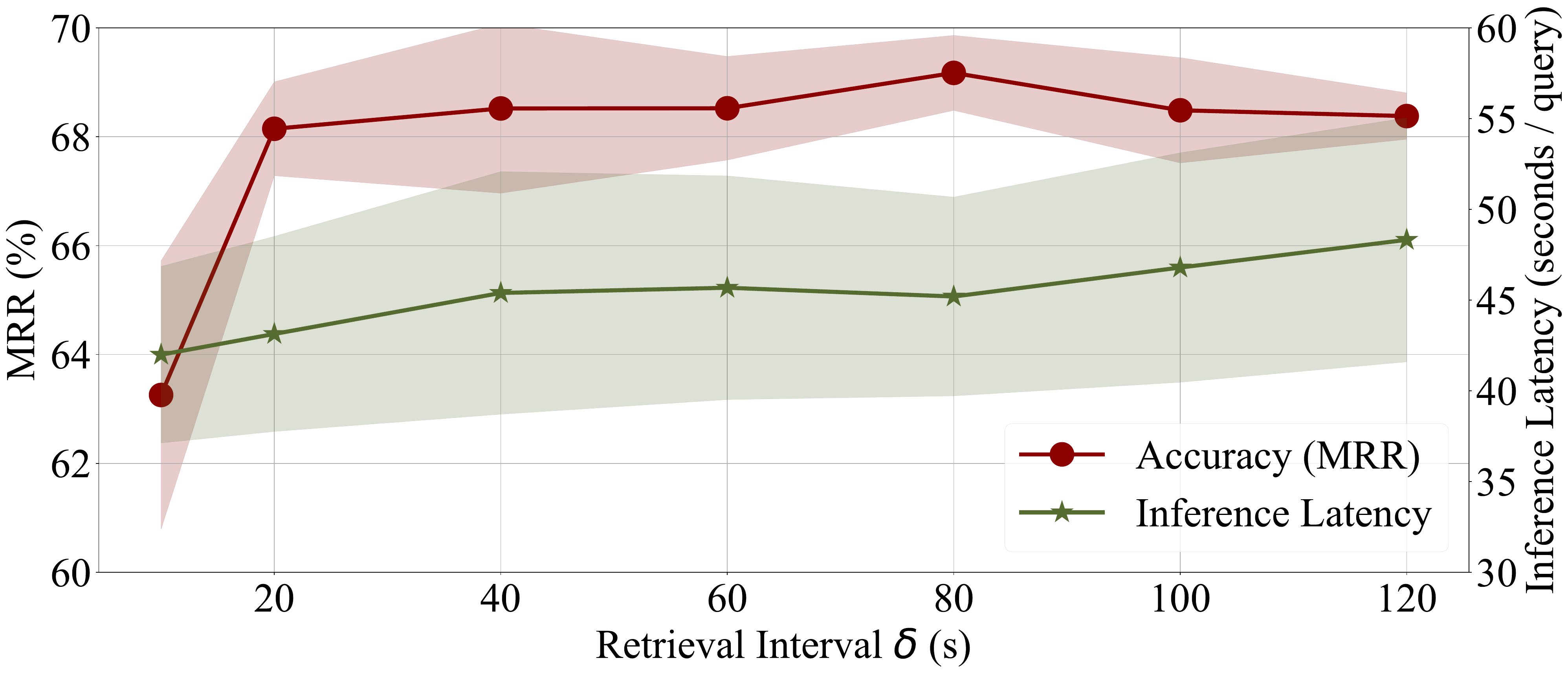}
	\caption{Hyperparameter analysis of the retrieval interval $\delta$ (s).}
	\label{fig: hyperparameter}
\end{figure}

As shown in Figure~\ref{fig: hyperparameter}, we conduct the hyperparameter analysis on the AIOps 2022-$\mathbf{A}$ dataset, reporting both localization accuracy (MRR) and inference latency. When the retrieval interval $\delta$ is small, the performance is suboptimal, as insufficient metric context is available to support reliable diagnosis. As $\delta$ increases, the accuracy improves and stabilizes, while further enlarging the interval leads to diminishing returns and slight performance degradation.

Meanwhile, inference latency increases monotonically with larger $\delta$. This is because a longer retrieval interval results in more retrieved metric data, which expands the input context provided to the LLM and consequently increases inference time. Overall, the results demonstrate a clear trade-off between accuracy and efficiency, and the relatively small variance across runs indicates that RCLAgent is robust to stochastic LLM outputs under different hyperparameter settings.

\section{Threats to Validity and Limitations}

Despite the promising results demonstrated by RCLAgent, several threats to validity should be acknowledged. First, the implementation and configuration of baseline approaches may introduce bias. For CRISP, MicroRank, and mABC, we directly used their publicly available source code. For other state-of-the-art baselines, we reimplemented the methods based on the descriptions provided in their respective papers. Although we carefully tuned hyperparameters and selected the best-performing configurations for all baselines through empirical experimentation, discrepancies between our implementations and the original authors’ setups may still affect the comparison results. Second, the evaluation relies on three benchmarks—AIOps 2022, Augmented-TrainTicket, and RCAEval—which, while extensive and derived from real-world microservice systems, may not fully capture the diversity of failure patterns, system scales, and operational environments encountered in practice. As a result, the generalizability of our findings to other domains or system architectures may be limited.

In addition, RCLAgent has several inherent limitations. 

(1) RCLAgent follows a trace-centric but not trace-only RCL process. In RCLAgent, distributed traces provide the structural backbone for organizing recursive multi-agent exploration along the invocation graph, while logs and metrics serve as complementary evidence sources for self-state verification and final diagnosis. Therefore, RCLAgent is most suitable when the target system has sufficient tracing coverage and when the failure propagation path can be reasonably reflected by the trace graph. When trace data are unavailable, the recursive multi-agent exploration cannot be naturally organized along the invocation graph. In such cases, RCLAgent would degenerate into a more conventional ReAct-style agent that repeatedly queries metrics and logs without explicit graph guidance. This would substantially increase the amount of data that a single LLM needs to retrieve and analyze, thereby increasing inference cost and latency while potentially reducing RCL accuracy.

(2) RCLAgent may be affected by blind-spot services in the trace graph. If some services are not instrumented or their spans are missing due to incomplete tracing, sampling, or observability failures, the constructed trace graph may not faithfully reflect the actual failure propagation path. Since RCLAgent performs recursive exploration based on the observed trace topology, missing nodes or edges may prevent the agent from reaching the true faulty component. In such scenarios, RCLAgent may either stop at an upstream observable symptom or misattribute the root cause to a neighboring service with available telemetry. More generally, the trace-centric process may also be less effective when important asynchronous dependencies are not captured by traces, or when failures do not manifest clearly through request-level propagation. Therefore, the effectiveness of RCLAgent depends on the completeness and quality of distributed tracing.

(3) RCLAgent inherits the limitations of LLM-based systems. Although the proposed framework constrains each agent with structured prompts, bounded context, and tool-provided evidence, the reasoning process may still be affected by hallucination, inconsistent judgments, or sensitivity to prompt design and model capability. Moreover, LLM inference introduces additional monetary cost and runtime overhead compared with lightweight statistical or graph-based methods. While RCLAgent mitigates this issue through parallel reasoning and bounded local contexts, deploying it in large-scale production environments still requires careful cost control, model selection, and concurrency management.

(4) RCLAgent relies on the reasoning capability of the underlying LLM, especially during Evidence Consolidation, where a Dedicated Agent needs to judge whether the available evidence supports attributing the root cause to itself, to one of its child agents, or to a higher-level aggregation process. We do not claim a formal guarantee that each local attribution decision is always correct. Instead, RCLAgent should be viewed as a structured framework that constrains and organizes LLM-based diagnosis through trace-aligned decomposition, bounded local contexts, structured evidence summaries, recursive aggregation, and the Global Evidence Graph. Therefore, RCLAgent is expected to work reliably under several practical preconditions: the trace graph should sufficiently cover the failure propagation path; logs and metrics should contain meaningful signals around the failure timestamp; and the LLM backbone should have adequate instruction-following, tool-use, and reasoning capability. As shown in Table~\ref{tab: llm-impact}, performance varies across different LLM backbones, confirming that model capability and task alignment remain important factors. When these preconditions are not satisfied, RCLAgent may produce low-quality local hypotheses or propagate incorrect evidence, which can affect the final diagnosis despite the use of bounded contexts and the Global Evidence Graph.

\section{Related Work}

\subsection{Trace-Based Root Cause Localization}

With the advancement of distributed tracing and its supporting infrastructure, researchers have explored various trace-based techniques for root cause localization. Several studies~\cite{li2022enjoy, luo2021characterizing, zhou2018fault, liu2025ora, he2025walk, he2025united, huang2025uda, liu2025aaad, duan2025logaction, xiao2025coorlog} have analyzed the role of distributed tracing in diagnosing failures in large-scale microservices systems, highlighting the importance of automated trace analysis for effective root cause localization.

Recently, machine learning techniques have been widely adopted for trace-based root cause localization. For instance, Zhou et al.~\cite{zhou2019latent} proposed MEPFL, a supervised learning-based approach. Gan et al.~\cite{gan2019seer} introduced Seer, a model leveraging CNNs and LSTMs for root cause analysis. Liu et al.~\cite{liu2020unsupervised} proposed TraceAnomaly, an unsupervised approach for anomaly detection in traces. Additionally, Gan et al.~\cite{gan2021sage} developed Sage, which utilizes graph neural networks for root cause localization.

Beyond machine learning-based methods, some researchers have extended spectrum analysis techniques to improve the practicality of trace-based root cause localization. For example, Yu et al.~\cite{yu2021microrank} proposed MicroRank, which was later extended into TraceRank~\cite{yu2023tracerank} by integrating spectrum analysis with a random walk-based method. Li et al.~\cite{li2021practical} combined spectrum analysis with frequent pattern mining to identify root cause service instances. Zhang et al.~\cite{zhang2024trace} introduced TraceConstract, which leverages sequence representations along with contrast sequential pattern mining and spectrum analysis to efficiently localize multi-dimensional root causes.

\subsection{LLM-based Failure Management}

Large language models, with their semantic understanding and logical reasoning capabilities, have significantly improved the field of failure management~\cite{zhang2025survey} and are increasingly becoming a focal point of research. Numerous LLM-based approaches have been proposed to address various aspects of failure management, including anomaly detection, failure diagnosis, and automated mitigation~\cite{rasul2023lag, liu2024timer, das2024decoder, shi2023shellgpt, liu2024anomalyllm, liu2024unitime, guo2023owl, liu2024loglm, chen2024automatic, jiang2024xpert, zhang2024lm, hamadanian2023holistic, pan2024raglog, zhang2024lograg, zhang2025xraglog, zhang2025scalalog, zhang2025agentfm, zhang2025thinkfl, zhang2025logdb, eagerlog, midlog, famos, logcae, llmelog, afalog, aclog, zhang2025surveyparallel, zhang2025adaptive, zhang2025microremed, pan2025omni, pan2025d, hong2025cslparser, zhang2026hypothesize, zhang2026runtimeslicer, zhang2026efficient, zhang2026e2e, zhang2026towards}.

Some studies have developed foundation models specifically for failure management. For example, Lag-Llama~\cite{rasul2023lag}, Timer~\cite{liu2024timer}, and TimesFM~\cite{das2024decoder} pretrain foundation models for metrics-based anomaly detection. Similarly, ShellGPT~\cite{shi2023shellgpt} trains a model capable of automatically generating shell scripts for automated mitigation.

Other approaches adopt fine-tuning strategies to tailor LLMs for failure management tasks. For instance, AnomalyLLM~\cite{liu2024anomalyllm} and UniTime~\cite{liu2024unitime} employ full fine-tuning for anomaly detection, while OWL~\cite{guo2023owl} and LogLM~\cite{liu2024loglm} leverage parameter-efficient fine-tuning techniques to build log analysis models.

Since these fine-tuning approaches require significant computational resources and time, an increasing number of methods rely on prompt-based techniques. For example, RCACopilot~\cite{chen2024automatic} and Xpert~\cite{jiang2024xpert} utilize in-context learning (ICL) to structure diagnostic processes, ensuring accurate root cause analysis. LM-PACE~\cite{zhang2024lm} applies chain-of-thought (CoT) reasoning to enhance GPT-4’s ability to analyze incident reports, while Hamadanian et al.~\cite{hamadanian2023holistic} extend this approach to generate mitigation solutions from incident reports. Additionally, RAGLog~\cite{pan2024raglog} and LogRAG~\cite{zhang2024lograg} use retrieval-augmented generation (RAG) to enhance log-based anomaly detection through historical log retrieval.

\section{Conclusion}

In this paper, we present RCLAgent, an in-depth root cause localization framework for microservice systems that realizes multi-agent recursion-of-thought with parallel reasoning. RCLAgent is motivated by our empirical study of how human SREs conduct root cause analysis in practice, as well as an analysis of why existing LLM-based approaches often fall short. By decomposing the diagnostic process along the trace graph, RCLAgent assigns each span to a Dedicated Agent operating within a bounded context, enabling focused and fine-grained analysis while selectively incorporating relevant log and metric evidence. These agents reason in parallel and propagate structured evidences upward, allowing the framework to balance deep localized reasoning with scalable global inference. The final diagnosis is produced by synthesizing the Root-Level Diagnosis Report and the Global Evidence Graph into a ranked list of root cause candidates with supporting rationales. Extensive experiments on multiple benchmarks demonstrate that RCLAgent outperforms state-of-the-art methods in both localization accuracy and inference efficiency.

In future work, we will further explore how to achieve more accurate and efficient root cause localization using smaller-scale models. Additionally, we are considering extending our approach to cover the entire failure management process.

\section*{Acknowledgment}

This work was supported by the Huawei–Peking University Joint Laboratory of Mathematics.

\balance
\bibliographystyle{IEEEtran}
\bibliography{mylib}

\newpage
\appendix

\subsection{Comparison with Tree- and Graph-Based LLM Approaches}

We further clarify the relationship between RCLAgent and existing tree- or graph-based LLM reasoning paradigms, including general reasoning frameworks such as Tree-of-Thought (ToT) and Graph-of-Thought (GoT), as well as recent LLM-based RCA systems such as mABC~\cite{zhang2024mabc}, RCAgent~\cite{wang2024rcagent}, and GALA~\cite{tian2025gala}, KnowledgeMind~\cite{ren2025multi} and Flow-of-Action~\cite{pei2025flow}. Although these approaches all introduce explicit structures or multiple agents to improve LLM reasoning, RCLAgent differs substantially in what the structure represents, how agents are organized, and how reasoning is executed for root cause localization.

In ToT- and GoT-style approaches, the tree or graph usually represents the reasoning process of a \emph{single LLM agent}. Nodes correspond to intermediate thoughts, candidate solutions, sub-steps, or partial conclusions, while edges represent logical dependencies among these reasoning states. Although such structures improve the organization of reasoning compared with a linear chain-of-thought, they still describe a single global inference trajectory. In other words, the structure mainly determines \emph{how one agent reasons}. As the reasoning tree or graph becomes deeper or broader, the agent must generate, evaluate, and aggregate more intermediate reasoning states, which can introduce substantial inference latency and context-management overhead.

Recent RCA-oriented methods move closer to our problem setting, but they still follow different design principles. mABC~\cite{zhang2024mabc} adopts a multi-agent collaboration framework in which multiple LLM-based agents coordinate through structured interaction and blockchain-style voting. While this design improves collaboration and robustness, its agents are organized around collaboration roles rather than being recursively grounded in the runtime trace topology. As a result, diagnosis is not decomposed into bounded span-level contexts in the same way as RCLAgent. RCAgent~\cite{wang2024rcagent} follows a centralized ReAct-style reasoning-and-acting paradigm, where a controller agent iteratively invokes external tools and updates its diagnosis within a single global reasoning trajectory. Although this improves tool-augmented RCA, it still relies on serial exploration and accumulated global context. GALA~\cite{tian2025gala} combines multimodal evidence with statistical causal guidance to improve RCA. However, its reasoning process is still not explicitly decomposed along the trace graph into recursively coordinated local diagnostic agents, making it fundamentally different from RCLAgent’s trace-aligned multi-agent recursion.

KnowledgeMind~\cite{ren2025multi} formulates fault localization as a Monte Carlo Tree Search process over a Fault Mining Tree, where metric, log, and trace agents provide evidence and a verifier agent assigns rewards to guide search. This design effectively constrains the search space, but its tree mainly serves as a sequential search structure for candidate service paths, rather than as a distributed execution structure in which multiple span-level agents reason concurrently. Flow-of-Action~\cite{pei2025flow} addresses RCA from another perspective by enhancing a ReAct-style agent with Standard Operating Procedures (SOPs), an action-set mechanism, and several auxiliary agents. While this improves the controllability of the RCA process through expert-defined procedural knowledge, the diagnosis remains centrally orchestrated by a MainAgent, and the auxiliary agents mainly support action planning and information filtering rather than localized diagnosis along trace branches.

In contrast, RCLAgent adopts a Multi-Agent Recursion-of-Thought paradigm, where the graph is not a reasoning graph constructed inside one agent's prompt, nor merely a search tree for sequential exploration. Instead, it is the actual trace graph of a microservice request. Each span in this trace graph is assigned to a Dedicated Agent, and each Dedicated Agent performs an independent, localized diagnostic process for its corresponding span. Specifically, each agent executes its own ReAct-style reasoning loop and selectively invokes data tools to examine the behavioral evidence associated with that span. Therefore, the nodes in RCLAgent do not represent intermediate thoughts of a single agent or candidate states in a search process; they represent distributed diagnostic agents grounded in concrete runtime entities.

This distinction leads to a different execution model. In ToT/GoT-style reasoning, dependencies among reasoning nodes often form a centralized reasoning process. RCAgent follows a centralized ReAct-style reasoning loop, in which a controller agent iteratively invokes tools and updates its diagnosis within a single accumulated context. mABC introduces multiple agents, but these agents are organized around collaboration roles and voting rather than being recursively grounded in the trace topology. GALA combines multimodal evidence with statistical causal guidance, but its reasoning process is still not decomposed into trace-aligned local diagnostic agents. In KnowledgeMind, MCTS iteratively explores and evaluates candidate service paths. In Flow-of-Action, multiple auxiliary agents assist a central MainAgent in following SOP-guided actions. By contrast, RCLAgent distributes reasoning responsibilities across trace-aligned Dedicated Agents. Agents corresponding to independent branches of the trace graph can reason concurrently, while parent agents only consolidate the distilled evidence reported by their child agents. Thus, reasoning in RCLAgent occurs \emph{across agents}, rather than merely \emph{within a graph-structured prompt}, a centralized ReAct loop, a role-based collaboration framework, a search tree, or a centrally controlled SOP flow.

This design is particularly important for root cause localization in microservice systems. Real-world traces are naturally hierarchical and branching: an abnormal request may propagate through many services, pods, and downstream calls. A monolithic or centrally aggregated LLM reasoning process must either place a large amount of heterogeneous runtime evidence into one context or repeatedly expand the reasoning/search structure as diagnosis proceeds. Both choices can lead to context explosion and increasing inference latency. RCLAgent mitigates this issue by bounding each agent's context to span-level evidence and by propagating only compact diagnostic summaries upward through Evidence Consolidation. Meanwhile, the Global Evidence Graph preserves fine-grained local hypotheses for final synthesis, reducing information loss during aggregation. These differences also help explain why RCLAgent achieves stronger accuracy and efficiency than baselines such as RCAgent, mABC, and GALA in our experiments.

In summary, ToT and GoT mainly structure \emph{how a single agent reasons}; RCAgent centralizes \emph{how a single controller explores with tools}; mABC organizes \emph{how multiple agents collaborate}; GALA strengthens \emph{how multimodal evidence is integrated}; KnowledgeMind guides \emph{which service path to search} through MCTS and knowledge-base rewards; and Flow-of-Action controls \emph{which diagnostic actions to take} through SOP-guided orchestration. RCLAgent instead redefines \emph{who reasons} by assigning diagnostic responsibilities to multiple span-level agents and coordinating them through the trace graph. This makes RCLAgent a distributed, domain-grounded, and parallelizable multi-agent framework specifically designed for in-depth root cause localization in microservice systems.

\subsection{Beyond Root Cause Localization}

Although the ultimate output of RCLAgent is a ranked list of root causes—a coarse-grained RCA—the reasoning process itself is fully interpretable and recorded. In particular, intermediate artifacts such as the Root-Level Diagnosis Report and the Global Evidence Graph capture step-by-step LLM outputs, with each step prompted to provide its reasoning. As a result, RCLAgent can also support fine-grained RCA, producing detailed root-cause indicators. These include anomalous metrics that reflect the failure, trace paths that reveal the propagation of faults, and summarized log entries that provide additional evidence of the underlying issue.

\subsection{Why LLM-based RCL?}
\label{sec:why-llm}

Although some non-LLM-based methods achieve strong accuracy on specific datasets, their performance often does not generalize well across different environments. For example, Nezha achieves an MRR that is 3.9\% higher than RCLAgent with Qwen-3.6-Plus on the Augmented-TrainTicket dataset. However, its performance drops substantially on AIOPS 2022 and RCAEval. This suggests that Nezha is highly dependent on dataset-specific assumptions and data relationships.

The main reason is that Nezha relies heavily on structural changes in trace graphs and specific correlations among logs, metrics, and traces. However, in RCAEval, the fault types are mainly resource-related faults, such as CPU, memory, disk, delay, and socket faults. These faults often affect runtime behavior but do not necessarily change the trace graph structure. As a result, methods that rely primarily on trace-structure variation may fail to capture the true root cause. Similarly, on AIOPS 2022, Nezha performs poorly for two reasons. First, the log modality is unavailable because AIOPS logs do not provide per-span SpanID linkage. Second, most faults in AIOPS 2022 are also resource-related faults that do not trigger span disappearance or obvious structural changes in the trace graph. In contrast, LLM-based RCL methods can flexibly integrate heterogeneous evidence, including traces, metrics, logs, and contextual descriptions, without being tightly coupled to a specific graph-change pattern. Therefore, LLM-based RCL provides stronger generalization across datasets and deployment environments.

From the efficiency perspective, LLM-based RCL methods are admittedly slower than many traditional non-LLM baselines, even though RCLAgent accelerates the diagnostic process through parallel multi-agent reasoning. This overhead mainly comes from the response latency of large language models. However, the runtime latency during diagnosis is only one part of the overall efficiency cost. Methods such as Nezha require substantial preparation before deployment. Specifically, they usually involve three environment-specific preprocessing steps. First, log template mining must be performed, where tools such as Drain3 are used to learn log templates from historical logs of the target system. Second, normal pattern construction is required, which scans weeks or months of historical trace data to build normal invocation graph structures. In our experiments, a single construction process needs to process approximately 9 million trace records. Third, metric thresholds must be generated by computing the historical mean and variance for each pod, which also requires sufficient historical metric collection. These steps must be repeated for every new deployment environment, introducing considerable upfront efficiency costs that are often overlooked in per-query latency comparisons.

Moreover, although efficiency is important for root cause localization, RCLAgent reduces the diagnosis time to within one minute in our experiments, which is already practical for many real-world incident response scenarios. In comparison, manual diagnosis by human SREs often takes more than ten minutes. More importantly, RCLAgent not only outputs a ranked list of root cause candidates, but also provides a structured reasoning process with supporting evidence. This makes it useful as an assistant for human SREs, helping them verify the diagnosis, understand the failure propagation path, and perform subsequent remediation more efficiently.

\end{document}